\documentstyle[prl,aps,amssymb,psfig]{revtex}

\begin{document}
\title{ Periodic Orbit Theory  and Spectral Statistics for Quantum Graphs}
\author{Tsampikos Kottos and Uzy Smilansky}
\address{{Department of Physics of Complex Systems,} \\
{The Weizmann Institute of Science, Rehovot 76100, Israel}}

\date{\today }
\maketitle

\begin{abstract}
We quantize graphs (networks) which consist of a finite number of bonds and
vertices. We show that the spectral statistics of fully connected graphs is
well reproduced by random matrix theory. We also define a classical phase
space for the graphs, where the dynamics is mixing and the periodic orbits
 proliferate exponentially. An exact trace formula for
the quantum spectrum is developed in terms of the same periodic orbits, and
it is used to investigate the origin of the connection between random matrix
theory and the underlying chaotic classical dynamics. Being an exact theory,
and due to its relative simplicity, it offers new insights into this problem
which is at the fore-front of the research in Quantum Chaos and related
 fields.\\
\end{abstract}
\hspace {1.5 cm} PACS numbers: 05.45.+b, 03.65.Sq

\vspace{0.8cm}
\hspace{4.5cm} Accepted for publication in the Annals of Physics



\section{\bf Introduction}

Quantum  graphs (networks) of one-dimensional wires connected at nodes were
introduced already more than half a century ago to model physical systems.
To the best of our knowledge, they appeared for the first time in connection
with free electron models of organic molecules
 \cite{P36,K48,P49,RS53,C54,M70,C71,RB72}.
The molecules were visualized as a set of atoms at fixed locations
connected by bond paths, along which the electrons obey a one-dimensional
Schr\"{o}dinger equation with an appropriate potential. In recent years the
interest in quantum graphs has been revived in many areas of physics,
and in particular in the context of condensed matter physics. Amongst the
 systems
which were successfully modeled by quantum graphs we mention e.g.,
studies of super-conductivity in granular and artificial materials \cite{A83},
single-mode acoustic and electro-magnetic waveguide networks \cite{FJK87,ML77},
Anderson transition \cite{A81} and quantum Hall systems with long range
potential \cite{CC88},  fracton excitations in fractal structures
\cite{NYO94}, and mesoscopic quantum systems \cite{Ibook}.
The construction of the wave equations for
such networks is a topic in its own right. Ruedenberg and Scherr \cite{RS53}
(see also \cite{RB72}), who were apparently among the first to address the
problem, based their formulation on the analysis of the limit of wires of
finite thickness.
Quantum graphs can be considered as idealizations of physical networks
in the limit where the widths of the wires are  much smaller than  all the
other length scales in the problem. Thus, neglecting the lateral size of the
wire, i.e., assuming that the propagating waves remain in a single transverse
mode, one replaces the corresponding partial differential Schr\"{o}dinger
equation by an ordinary differential operator. This can be justified assuming
that the inter-mode coupling involves a dynamical tunneling and therefore it
diminishes exponentially with the decreasing wire thickness. Moreover, when
no external field is applied, the motion on the bonds is free, and the problem
can be further reduced to finite matrices \cite{A83,A94}. Alexander \cite{A83}
was probably the first to discuss networks in external magnetic fields.

Quantum  graphs  attracted the attention of the mathematics community as
well. J. P. Roth \cite{R83} was probably the first to derive  a trace formula
for the spectrum of a Laplacian on graphs. Recently the problem came
of age in a series of mathematical works by Exner and Seba \cite{ES89,EG96},
Avron \cite{A94,AEL94} and Carlson \cite{Cpre}, whose formulation is based
on the von-Neumann theory of self-adjoint extensions of formal differential
operators (see also \cite{KSpre} and references therein).

In spite of all this activity, the statistical properties of the spectra
displayed by quantum
graphs were hardly investigated in the past \cite{CC88,KS97}. Our
motivation for
studying these spectral properties comes from the theory of Quantum Chaos
which deals with quantum systems exhibiting chaotic motion in the classical
limit. One of the main observations of this field is that in the extreme case,
when classical motion is strongly chaotic, and in the limit $\hbar
\rightarrow 0$, the statistical properties of spectra are well described by
 Random Matrix Theory (RMT). At the same time, the spectra of
quantized integrable systems display Poissonian statistics.
An important goal of quantum chaology, is to develop a theory, which relates
the quantum spectral statistics to the underlying features of classical
dynamics. The main tool in this endeavour consists of trace formulae which
provide an expression for the spectral density in terms of classical periodic
manifolds - isolated orbits for chaotic systems, and tori for integrable ones.
In most cases, only the semiclassical approximation for the trace formulae
are known \cite{G90} and their application is not only hampered by the
intrinsic complexity of the set of periodic orbits, but also by the doubts
about the ability of the semiclassical trace formulae to provide accurate
enough basis for the further developments. This motivated the introduction
and the  study of particular ``toy" systems where the required periodic-orbit
data can be easily accumulated, while at the same time the trace formulae are
exact rather than semi-classical \cite{K91}. Unfortunately, only very few
models combine the desirable features of both behaving ``typically'' and
being mathematically simple. It is the main purpose of this paper to propose
quantum graphs as a very convenient and rich class of systems where the  above
mentioned requirements are met satisfactorily. We shall show that for quantum
graphs, one can write an exact trace formula, which is based on ``periodic
orbits" in a way which is analogous to the known trace formulae for chaotic
systems. Moreover, we shall define the corresponding underlying classical
dynamics, and write down the relevant Frobenius Perron operator for the
``phase space" evolution. This analogy will enable us to study further the
connection between spectral fluctuations and the classical dynamics. Another
clear advantage of quantum graphs is the relative ease by which a large number
of spectral data can be computed. This enables a rather accurate numerical
studies of systems and problems for which  analytical results are lacking or
insufficient.

 This paper  extends our  previous report \cite{KS97} on the spectral
properties of quantum graphs, both in detail and depth. In particular,
we address issues which have to do with the transition which the
spectral statistics undergo when the connectivity of the graph is altered.
This problem is intimately connected with the semiclassical theory of
Anderson localization. We also add another parameter to the model, which
enables us to interpolate between Dirichlet and Neumann boundary conditions.
We show that this induces a transition between integrable and chaotic
dynamics, and
we study its effect on the spectral statistics.

The paper is structured in the following way. In Sec.~II, the mathematical
model is introduced and the main definitions are given. In Sec.~III we are
describing three methods of quantizing graphs. The first method is rather
standard  (see e.g.,\cite {A94}) and it is the most convenient
  for numerical computations.  The two  other methods  are
related to the scattering approach to quantization \cite {S94}. One of them
forms the basis for the development of the exact trace formula, and it
singles out a unitary  matrix of dimension $ 2B \times  2B $
where $B$ is the number of bonds on the graph. This matrix
is  the main building block of our theory, and we refer to it
as the ``bond scattering matrix" $S_B$. An alternative quantization condition
is achieved in terms of the  ``vertex
scattering matrix" $S_V$. It is of dimension $V\times V$ where $V$ is the
number of
vertices, and it describes the transport through a system in which each
of the vertices is attached to a conducting wire. In Section IV we
present the trace formula for the quantum graph, and also
  express  the spectral $\zeta$ function as a sum over composite
periodic orbits. Section IV terminates with the introduction
and the discussion of the underlying classical system. In Section V, the
statistical properties  of the eigenphase spectrum of the bond
scattering matrix $S_B$
and of the energy (or wavenumber) spectrum are analyzed and compared with the
predictions of RMT
and of periodic orbits theory. In Section VI, we analyze two families of graphs
which are not uniformly connected. The resulting spectral statistics deviate
from the expectations of RMT, and we
explain these deviations using  periodic orbit theory.
 Within this study, we investigate also the
localization/delocalization transition experienced by graphs as a
function of the connectivity. Our conclusions are summarized in the last
section (Section VII).


\section{\bf Quantum Graphs: Definitions}

In this section we shall present and discuss the Schr\"odinger operator for
graphs. We start with a few definitions. Graphs consist of $V$ {\it vertices}
connected by $B$ {\it bonds} (or {\it edges}). The {\it valency}
$v_{i\text{ }}$
of a vertex $i$ is the number of bonds meeting at that vertex. The graph is
called $v${\it -regular} if all the vertices have the same valency $v$. When
the vertices $i$ and $j$ are connected, we denote the connecting bond by
 $b=(i,j)$.
The same bond can also be referred to as $\vec{b} \equiv
(Min(i,j),Max(i,j))$ or
$\smash{\stackrel{\leftarrow}{b}} \equiv (Max(i,j),Min(i,j))$ whenever we need
 to assign a
direction to the bond. Several bonds connecting the same two vertices are
called multiple bonds and the corresponding graph is called a multi-graph.
Finally, a bond with coinciding endpoints is called a {\it loop}. In what
follows, unless explicitly specified, we shall  consider graphs
without multiple bonds or loops. Moreover, we shall treat only connected
graphs.

Associated to every graph is its {\it connectivity (adjacency) matrix}
 $C_{i,j}$.
It is a square matrix of size $V$ whose matrix elements $C_{i,j}$ are given in
the following way
\begin{equation}
C_{i,j}=C_{j,i}=\left\{
\begin{array}{l}
1\text{ if }i,j\text{ are connected} \\
0\text{ otherwise}
\end{array}
\right\} ,\bigskip\ i,j=1,...,V.  \label{cmat}
\end{equation}
For loop-less graphs the diagonal elements of $C$ are zero.
The connectivity matrix of connected graphs cannot be written
as a block diagonal matrix.
The valency of a
vertex is given in terms of the connectivity matrix, by $v_i= \sum_{j=1}^V
C_{i,j}$ and the total number of  bonds is $B= {1\over
2}\sum_{i,j=1}^VC_{i,j}$.
 As will be shown bellow (Sec. IV.C), the topological characterization of
the graph which was given above, is sufficient for the study of ``classical
dynamics'' on graphs.

For the quantum description we assign to each bond $b=(i,j)$ a coordinate
$x_{i,j}$ which indicates the position along the bond. $x_{i,j}$ takes the
value $0$ at the vertex $i$ and the value $L_{i,j} \equiv L_{j,i}$ at the
vertex $j$ while $x_{j,i}$ is zero at $j$ and $L_{i,j}$ at $i$. We have thus
defined the {\it length matrix} $L_{i,j}$ with matrix elements different from
zero, whenever $C_{i,j}\neq 0$ and $L_{i,j}=L_{j,i}$ for $b=1,...,B$. The
derivations presented in the sequel are valid for any choice of the lengths
$L_{i,j}$. However, in some applications we would avoid non generic
degeneracies
by assuming that the $L_{i,j}$ are {\it rationally independent}. The mean
length
is defined by $\left \langle L \right \rangle \equiv {1\over B}\sum_{b=1}^B
 L_b$.

The wavefunction $\Psi$ is a $B-$component vector and will be written as
$\left(\Psi_{b_1}(x_{b_1}),\Psi_{b_2}(x_{b_2}),...,\Psi_{b_B}(x_{b_B})\right)^T$
where the set $\{b_i\}_{i=1}^B$ consists of $B$ different bonds. We will call
$\Psi_b(x_b)$ the component of $\Psi$ on the bond $b$. The bond coordinates
$x_b$ were defined above. When there is no danger of confusion, we shall use
the shorthand notation $\Psi_b(x)$ for $\Psi_b(x_b)$ and it is understood that
$x$ is the coordinate on the bond $b$ to which the component $\Psi_b$ refers.

The Schr\"{o}dinger operator (with $\hbar = 2m = 1$) is defined on the graph
in the following way \cite{A83,A94}: On each bond $b$, the component
$\Psi_b$ of
the total wave function $\Psi$ is a solution of the one - dimensional equation
\begin{equation}
\left( -i{\frac{{\rm d\ \ }}{{\rm d}x}}-A_b\right) ^2\Psi _b(x)=k^2\Psi
_b(x), \,\,\,\,\,\,\,\,\,\,\,\, \bigskip\ b=(i,j)  \label{schrodinger}.
\end{equation}
We included a ``magnetic vector potential" $A_b$ (with $\Re e(A_{b})\ne 0$
and $A_{\vec{b}}= -A_{_{\smash{\stackrel{\leftarrow}{b}}}}$) which breaks
the time reversal symmetry. In most applications we shall assume that all
the $A_{b}$'s are equal and the bond index will be dropped.

The wave function must satisfy boundary conditions at the vertices, which
ensure continuity (uniqueness) and current conservation. The imposition of
these
boundary conditions guarantees that the resulting Schr\"odinger operator is
self-adjoint. The continuity condition requires that at each vertex $i$, the
wave function assume a value denoted by $\varphi_i$ which is independent of the
bond from where the vertex is approached. Current conservation imposes a
 condition
on the derivatives of the wave function at the vertices. By assuming that
$\{b_i\}_{i=1}^B = \{\vec{b}_i\}_{i=1}^B$ the conditions are explicitly
 specified
in the following way. For every $ i=1,\cdots, V$:
\begin{equation}
\left\{
\begin{array}{l}
\bullet {\rm Continuity: }
 \\
\bigskip\ \Psi _{i,j }(x)|_{x = 0} = \varphi _i ,\,\,\,\,\,\,\,\,\,\,
\Psi _{i,j}(x)|_{x = L_{i,j}}=\varphi_j \ \ {\rm for \ all } \ \ i<j \ {\rm
and} \ C_{i,j} \ne 0\\
\bullet \text{{\rm Current conservation:}}
 \\
\bigskip\ \sum_{j<i} C_{i,j}\left( iA_{j,i}-{\frac{{\rm d}\ \ }{{\rm
 d}x}}\right)
\Psi _{j,i}(x)|_{x= L_{i,j}} + \sum_{j>i} C_{i,j}\left( -iA_{i,j} +
{\frac{{\rm d}\ \ }{{\rm d}x}}\right) \Psi _{i,j}(x)|_{x=0}=\lambda
_i\varphi _i
 \ .
\end{array}
\right.  \label{current}
\end{equation}
  The parameters $\lambda _i$ are free parameters which
determine the boundary conditions. In many applications we shall assume
that the $\lambda _i$ are all equal, and in such cases the vertex index
will be dropped.  In the  case when $v_i =2$, the matching conditions
can be represented  by a $\delta$-function potential of strength
$\lambda _i$. By analogy, we shall  refer to the $\lambda_i$  as the
{\it vertex scattering potential}. In the sequel, we shall always
assume that $\lambda_i \geq 0$ and will consider the  domain $k^2\geq 0$
(which excludes states bounded at a single vertex). The special
case of zero $\lambda_i$'s, corresponds to Neumann boundary conditions.
 Dirichlet boundary conditions are introduced when all the  $\lambda_i
=\infty $. This implies   $\varphi_i =0$ for all $i$, thus turning the
graph into a union of non interacting bonds. A finite value of $\lambda_i$
introduces a new length scale. It is natural therefore, to interpret it
in physical terms as a representation of a local impurity or an external
fields \cite{ES89,EG96,myfav}. We finally note that the above model can
be considered as a generalization of the Kronig-Penney model to a multiply
connected, yet  one dimensional manifold.


\section{\bf The Spectrum of Quantum Graphs}

The set of boundary conditions (\ref{current}), discussed in the previous
section, ensures that the Schr\"{o}dinger operator (\ref{schrodinger}) is
self-adjoint, and hence the existence of an unbounded, discrete spectrum
$\left\{ k_n^2\right\}$. In the
following three subsections we shall introduce three different approaches
which can be used for the calculation of the wavenumbers spectrum
$\left\{ k_n\right\}$. These
approaches complement each other and enable us to address various aspects
of the quantum graphs using different points of view.


\subsection{\bf The Vertex Secular Equation}

The eigenfunctions of the graph are completely determined by their values
at the vertices $\left \{ \varphi _i\right \}_{i=1}^V$. The quantization
condition which is to be derived here, specifies the values of $k$ for
which a non trivial set of $\left \{ \varphi _i\right \}_{i=1}^V$'s can
be found \cite{A83,A94}.

The wave function  $\Psi$ is constructed from  $B$ components which
correspond to the various bonds. At any bond $b=(i,j)$  the component
 $\Psi_b$ can be written in terms of its values on the vertices  $i$
and $j$ as
\begin{equation}
\Psi _{i,j}={\frac{e^{iA_{i,j}x}}{\sin kL_{i,j}}}\left( \varphi _i\sin \left[
k(L_{i,j}-x)\right] +\varphi _je^{-iA_{i,j}L_{i,j}}\sin kx\right)
 C_{i,j},\,\,\,\,\,
\,\,\,\, i<j. \label{wfun}
\end{equation}
$\Psi$ has, by construction, a unique value on the vertices
 and satisfies the Schr\"odinger equation ~(\ref{schrodinger}).
The current conservation condition (\ref{current}) leads to
\begin{eqnarray}
-\sum_{j< i}\frac{ke^{iA_{j,i}L_{i,j}}C_{i,j}}{\sin (kL_{i,j})}\left(
 -\varphi _j
+\varphi _ie^{-iA_{j,i}L_{i,j}}\cos (kL_{i,j})\right) & &\nonumber \\
+\sum_{j>i}\frac{kC_{i,j}}{\sin (kL_{i,j})}\left( -\varphi
 _i\cos (kL_{i,j})+\varphi
_je^{-iA_{i,j}L_{i,j}}\right) &=&\lambda _i\varphi _i,\,\,\,\,\,\,\,\,\,
\bigskip\ \forall i.
\label{meq}
\end{eqnarray}
This is a set of linear homogeneous equations for the $\varphi $'s which
has a nontrivial solution when
\begin{equation}
\label{secu}
\det\left( h_{i,j}(k,A)\right) =0 \ ,
\end{equation}
where
\begin{equation}
\label{secu1}
h_{i,j}(k,A) =\left\{
\begin{array}{cc}
-\sum_{m\neq i}C_{i,m}\cot(kL_{i,m})-\frac{\lambda _i}k , &
i=j \\ \\
C_{i,j}e^{-iA_{i,j}L_{i,j}}(\sin (kL_{i,j}))^{-1}, &  i\neq j \ .
\end{array}
\right.   \nonumber
\end{equation}

The terms $h_{l,m}=h_{m,l}^*,\, h_{l,l}$ and $h_{m,m}$ in (\ref{secu1}) diverge
when $k$ is an integer multiple of ${{\pi }\over {L_{l,m}}}$ .
This can be easily rectified  by  replacing the diverging  terms by
\begin{eqnarray}
 \label{secus}
h_{l,m}(k,A)&=&h_{m,l}(k,A)=0 \\
h_{l,l}(k,A)&=&-\sum_{j^{\prime }\neq m}C_{l,j^{\prime }}
\cot(kL_{l,j^{\prime }})-\frac{
\lambda _l}k,\,\,  \nonumber \\
h_{m,m}(k,A)&=&-\sum_{j^{\prime }\neq l}C_{j^{\prime },m}\cot(kL_{j^{\prime
},m})- \frac{
\lambda _m}k.  \nonumber
\end{eqnarray}

The secular equations (\ref{secu}-\ref{secus}) for the quantized graph can be
solved numerically to provide an arbitrarily large sequence of eigenvalues
$\left\{ k_n\right\} $.

As we said previously, the effect of the Dirichlet boundary condition on
all the vertices  ($\lambda _i=\infty ,\,\,\,\forall i$) is to  disconnect
the bonds. The eigenfunctions in this case have a simple structure:
they vanish on all the bonds except on one of the  bonds, $b$, where
\begin{equation}
\Psi _b=\frac{e^{iA_bx}}{\sqrt{L_b}}\sin (\frac{n_b\pi x}{L_b}),\bigskip\
\,k_n^{(b)}=\frac{n_b\pi }{L_b}  \ \ \ \  {\rm for\  all\ \  } n_b >0  \ ,
\label{diri1}
\end{equation}
for all $b$. The spectrum is the union of the individual spectra, and when
the lengths $L_b$ are rationally independent, the resulting spectrum displays
some Poissonian features. When the $\lambda_i$ are large but finite there is
always a small probability of ``leaking" of the wave functions through the
vertex scattering potential. As $k$ increases, however, all intermediate
boundary conditions converge to the Neumann limit.


\subsection{\bf Scattering Approach - The bonds $S_B$ Matrix}

The quantization of graphs can be accomplished in a different way which
although less efficient from the numerical point of view, provides us
with a natural and convenient starting point for the construction of the
trace formula. It is an example of the scattering approach to quantization
\cite{S94}. Another variant of this method will be presented in the following
subsection.

We first introduce the  scattering matrix related to a single vertex. This
is done by solving an auxiliary problem of a single vertex $i$, say, with
$v_i$ emanating bonds which extend to infinity. The wave-function $\Psi^{(i)}$
has components on all the bonds $ b_j\equiv (i,j) , j=1,\cdots,v_i \ $, which
emerge from the $i$ vertex (note that we enumerate and denote the bonds in
the auxiliary problem by their analogues on the original graph). $\Psi^{(i)}$
can be written as a linear combination of functions $\Psi^{(i,j)}$ which are
solutions for the case where there is an incoming wave entering $i$ from $b_j=
(i,j)$ and outgoing waves from $i$ to all bonds $b_{j'}$ (including $j=j'$
which
correspond to the reflected part). $\Psi^{(i,j)}$ is a $v_i-$dimensional vector
with components $\Psi_{ j' }^{(i,j)} (x_{j'})$ for all $1\le j'\le v_i\ $,
\begin{equation}
\Psi_{ j' }^{(i,j)} (x_{j'})\ \ =  \delta_{j,j'} e^{-ikx_{j'}+iA_{i,j}x_{j'}}+
\sigma _{j,j'}^{(i)}
 e^{ikx_{j'}+iA_{i,j'}x_{j'}} \ .  \label{scat1}
\end{equation}
Here the $x_{j'}$ are the distances from the vertex $i$ along the bonds
$(i,j')$, and
$\sigma _{j,j'}^{(i)}$ is the $v_i\times v_i$
  scattering matrix, which provides a transformation between the incoming and
the outgoing waves at the vertex $i$. The matching conditions ~(\ref{current})
 at the vertex ($x_j=0$) together with  (\ref
{scat1}), can be used to determine
$\sigma _{j,j^{\prime }}^{(i)}$:
\begin{equation}
\sigma _{j,j^{\prime }}^{(i)}=\left( -\delta _{j,j^{\prime }}+{\frac{
(1+e^{-i\omega _i})}{v_i}}\right) C_{i,j}C_{i,j^{\prime }},\medskip\
\,\,\,\omega _i=2\arctan \frac{\lambda _i}{v_ik}  \ .
\label{smatrix}
\end{equation}
  For the Dirichlet boundary conditions we get $\sigma_{j,j^{\prime }} ^{(i)}=
-\delta _{j,j^{\prime }}$ which indicates total reflection. For the Neumann
boundary condition
$\sigma _{j,j^{\prime }}^{(i)}=-\delta _{j,j^{\prime }}+\frac 2{v_i}$
which is independent of $k$. For any intermediate boundary condition,
the parameter that controls the scattering
process is the $k$ dependent parameter
\begin{equation}
 \Lambda_i \equiv\frac{\lambda _i}{v_ik}\ .
\label{lambdadef}
\end{equation}
The scattering matrix approaches the Neumann expression as $k \rightarrow
\infty$.
 Note that in all the non trivial cases ($v_i> 2$), back-scattering
 ($j =j^{\prime}$) is singled
out both in sign and in magnitude: $\sigma _{j,j }^{(i)}$ has always a negative
real part,
and the reflection probability $|\sigma _{j,j }^{(i)}|^2 $ approaches
$1$ as   the valency $v_i$ increases. One can easily check that
$\sigma ^{(i)}$ is a symmetric unitary matrix, ensuring flux conservation and
time reversal symmetry at the vertex.  For the Neumann boundary conditions
$\sigma ^{(i)}$ is a real orthogonal matrix.

We now write the general expression for an eigenfunction of the quantum graph
in terms of its components on the bonds $b=(i,j)$. We write the {\it same} bond
wave function in two ways. First, we use the standard notation introduced
before and call it $\Psi _{i,j}(x_j)$ where $x_j$ is the distance  from $i$.
The second representation employs the ``time reversed" notation
where the wave function is denoted by $\Psi _{j,i}(x_i)$ and  $x_i$ is
the distance from $j$. The general expressions for
the wave function  in the two representations read
\begin{eqnarray}  \label{outin}
\Psi _{i,j}(x_j) &=&a_{i,j}e^{i(-k+A_{i,j})x_j}+b_{i,j}e^{i(k+A_{i,j})x_j}  \\
\Psi _{j,i}(x_i) &=&a_{j,i}e^{i(-k+A_{j,i})x_i}+b_{j,i}e^{i(k+A_{j,i})x_i}
 \nonumber \\
&=&a_{j,i}e^{i(k-A_{j,i})x_j}e^{i(-k+A_{j,i})L_{i,j}} +
 b_{j,i}e^{i(k+A_{j,i})L_{i,j}}e^{i(-k-A_{j,i})x_j}  \ .  \nonumber
\end{eqnarray}
The  two representations describe the same function. This gives,
\begin{equation}
b_{i,j}=a_{j,i}e^{-ikL_{i,j}-iA_{i,j}L_{i,j}},\,\,\,\,
\,b_{j,i}=a_{i,j}e^{-ikL_{i,j}-iA_{j,i}L_{i,j}} \  .  \label{outin1}
\end{equation}
In other words, but for a phase factor, the outgoing wave from the vertex
$i$ in the direction $j$ is identical to the incoming wave at $j$ coming
from $i$.  The incoming and outgoing components of the wavefunction impinging
on the $i$'th  vertex satisfy
\begin{equation}
b_{i,j}=\sum_{j^{\prime }}\sigma _{j,j^{\prime }}^{(i)}a_{i,j^{\prime }} \ .
\label{outin2}
\end{equation}
 ~(\ref{outin1}-\ref{outin2}) can be combined to a set of $2B$ homogeneous
linear equations for the coefficients $a_{i,j}$ which describe the wave
function
on each of the bonds. $a_{i,j}$ is the amplitude for propagation from $i$
to $j$
vertex along the bond $b=(i,j)$, while, $a_{j,i}$ is the amplitude for the
propagation in the ``time reversed" direction (i.e. from $j$ to $i$) along the
same bond. This distinction corresponds to assigning directions to the
bonds, so
that $\vec{b}$ and ${\smash{\stackrel{\leftarrow}{b}}}$ are considered as
 different entities.
We thus see that the present approach for quantizing the graph singles out the
description in terms of {\it directed bonds} as the natural setup.

The condition for a non trivial solution for the $2B$ dimensional vector
$(a_{\vec{b}_1}, \cdots,
 a_{\vec{b}_B},a_{_{\smash{\stackrel{\leftarrow}{b}_1}}}, \cdots,
a_{_{\smash{\stackrel{\leftarrow}{b}_B}}})^T$ gives the secular  equation for
 the total graph,
of the form \cite{Cpre}
\begin{equation}
\zeta_B(k)=\det \left[ I-S_B(k,A)\right] =0.  \label{secular}
\end{equation}
Here, the ``bond scattering matrix'' $S_B(k,A)=D(k;A)T$ is a unitary matrix
in the $2B$ dimensional space of directed bonds. It is a product of a
diagonal unitary matrix $D(k,A)$ which depends on the metric properties of
the graph, and a  unitary matrix $T$ which depends  on
the connectivity and on the vertex scattering potentials.
\begin{eqnarray}  \label{DandT}
D_{ij,i^{\prime }j^{\prime }}(k,A) &=&\delta _{i,i^{\prime }}\delta
_{j,j^{\prime }}{\rm e}^{ikL_{ij}+iA_{i,j}L_{ij}}\ ;{\rm with } \ L_{ij} =
L_{ji}
\ \ {\rm and }\ \  A_{ij} = -A_{ji} \  \label{scaco} \\
\ T_{ji,nm} &=&\delta _{n,i}C_{j,i}C_{i,m}\sigma _{ji,im}^{(i)}. \nonumber
\end{eqnarray}
The  matrix elements of $T$  assign  an  amplitude to a transition from one
directed bond to another. Such a transition can occur only if the
directed bonds are connected, that is, one is incoming and the other is
outgoing
from the same vertex. The phase and magnitude of the amplitude is given
by the corresponding matrix element of the single vertex scattering matrix.
>From (\ref{secular}) it follows that $k_n$ belongs to the wavenumber
spectrum  if and only if $S_B(k_n,A)$ has an
eigenvalue $+1$. As no approximations were made at any step of the
derivation, this quantization condition is exact. Furthermore, this gives a
constructive method to obtain not only the eigen-energies, but also the wave
function, in terms of the eigenvector of $S_B(k_n,A)$ with the eigenvalue $+1$
\cite{S94}.

The ``bond scattering matrix" $S_B$ cannot be associated with an actual
scattering system in the usual sense of scattering theory. Nevertheless we
shall keep referring to it as a scattering matrix, since it yields a
quantization condition which is of the standard form in the scattering
approach.

We finally comment that we use the letter $\zeta_B$ to
denote the secular function (\ref{secular}), because it can be
cast in a form which is reminiscent of the Riemann-Siegel expression for
the Riemann $\zeta$ function on the critical line. This will be shown
in chapter IV.

\subsection{\bf Scattering Approach - The vertex $S_V$ Matrix}

The vertex scattering matrix $S_V$ is obtained by converting the
graph of interest into a proper scattering system. This is done by attaching
 a lead which is extended to infinity  at each of the graph vertices.
A scattering solution with an incoming wave only in the lead $l$,
and outgoing waves on all the leads can be written in the following way.
On the external leads,
\begin{equation}
\Psi _i^{(l)}(x) = \delta _{i,l}e^{-ikx}+(S_V)_{i,l}e^{ikx} \  .
\end {equation}
On the $B$ internal bonds ,
\begin{equation}
\Psi _{i,j}^{(l)}(x) = {\frac{C_{i,j}e^{iA_{i,j}x}}{\sin kL_{i,j}}}\left(
\varphi
 _i^{(l)}\sin \left[
(L_{i,j}-x)k\right] +\varphi
 _j^{(l)}e^{-iA_{i,j}L_{i,j}}\sin kx\right),\,\,\,\,\,\,\, \bigskip
i<j.
 \label{phsca1}
\end{equation}

By applying the continuity and current conservation conditions (\ref{current})
at all the vertices, we get
\begin{eqnarray}
&&
\begin{array}{c}
\varphi_i^{(l)} = \delta_{i,l} + (S_V)_{i,l} \\
\left( -\delta_{i,l} +(S_V)_{i,l}  \right) i + \sum_{j=1}^{V}h_{i,j}(k)
\varphi_j^{(l)} = 0
\end{array}
  \label{phsca2} \\
&&  \nonumber
\end{eqnarray}
where $h(k)$ is the secular matrix defined in   (\ref{secu1},\ref{secus}).
Combining the above two equations we finally get for $S_V$
\begin{equation}
S_V = (i I + h(k))^{-1} (i I-h(k)).  \label{phsca3}
\end{equation}
where $I$ is the $V\times V$ unit matrix.

$S_V$ is unitary since $h(k)$ is hermitian, which ensures current conservation.
The graph spectrum can be identified as the set of wavenumbers for which $S_V$
has $1$ as an eigenvalue. This corresponds to a solution where no current flows
in the leads so that the continuity equations are satisfied on the internal
bonds
 (see \cite{S94}). $1$ is in the spectrum of $S_V$ if
\begin{equation}
\zeta_V(k)\equiv \det\left[ I -S_V\right] = 0 \longleftrightarrow 2^V
\det\left[ i I +
h(k)\right]^{-1} \det h(k)= 0  \label{phsca4}
\end{equation}
which is satisfied once $\det \,h(k) = 0$. This is identical with
the condition (\ref{secu}) which was derived in subsection III.A.

Variations on the same theme can be obtained by considering graph scattering
systems where leads are attached to an arbitrary  set of $L$ vertices
$\left \{i_l\right\}_{l=1}^L$, with $ 1 \le L < V$. The  $L \times L$
scattering
matrix $S_V$ has to be modified in the following way
\begin{equation}
S_V = 2i W \left ( h(k) + i W^{T} W \right ) ^{-1} W^{T} -I
\label{phsca5}
\end{equation}
where $W_{i_l,j}=\delta_{i_l,j}$ is the $L\times V$ leads - vertices coupling
matrix. In the case that we examined previously with $L=V$,  $W=I$.

The matrices $S_V$ will not be studied any further in this work,
and the derivation above was given for the sake of completeness. The $S_V$
scattering matrix corresponds to proper scattering problems, and can be
used to model experimental systems, such as e.g., conductance of mesoscopic
microdots. We shall study the  statistics of conductance fluctuations, based on
the vertex scattering matrices $S_V$ in a separate publication \cite{KS98}.

\section{\bf Periodic orbits, the trace formula and classical dynamics on
graphs}

In this section we derive an expression for the quantal density of states in
terms of periodic orbits on the graph. A trace formula for the Laplacian on
graphs was first presented by J. P. Roth \cite{R83}. Our result (see also
\cite{KS97}) generalizes Roth's expression in several ways, and it is derived
by means of a different approach. The key element in this theory is the concept
of a periodic orbit on the graph, which we shall introduce at this point.

An {\it orbit} on the graph is an itinerary (finite or infinite) of
successively
connected {\it directed} bonds $(i_1,i_2), (i_2,i_3),\cdots $. For graphs
without loops or multiple bonds, this is uniquely defined by the sequence of
vertices $i_1,i_2, \cdots$ with $i_m \in [1,V]$ and $C_{i_m,i_{m+1}} =1$
for all
 $m$.
An orbit is {\it periodic} with period $n$ if for all $k$, $
(i_{n+k},i_{n+k+1})
=(i_k,i_{k+1})$. The {\it code} of a periodic orbit of period $n$ is the
 sequence
of $n$ vertices $i_1,\cdots,i_n$ and the orbit consists of the bonds
$(i_m,i_{m+1})$ (with the identification $i_{m+n} \equiv i_{m}$). In this way,
any cyclic permutation of the code defines  the same periodic orbit.

The  periodic orbits (PO) can be classified in the following way:
\begin{itemize}
\item {\it Irreducible periodic orbits} -  PO's which do not intersect
themselves
so that any vertex label in the code can appear at most once. Since the graphs
are finite, the maximum period of irreducible PO's is $V$. To each
irreducible PO
corresponds its time reversed partner whose code is read in the reverse order.
The only code which is both irreducible and conjugate to itself under time
reversal is the code corresponding to PO's of period 2.
\item {\it Reducible periodic orbits} -  PO's whose code is constructed by
inserting the code of any number of irreducible PO's at any position which is
consistent with the connectivity matrix. All the PO's of period $n >V$
are reducible.
\item {\it Primitive periodic orbits} -  PO's whose code cannot be written down
as a repetition of a shorter code.
\end{itemize}

After these preliminaries, we are set to derive the trace formula for the
graphs. Once this is done, we shall show that one can define classical dynamics
on the graph, and that the periodic orbits on the graph are indeed the analogue
of the periodic orbits of hyperbolic classical Hamiltonian systems.


\subsection{\bf The Trace Formula}

The starting point for the derivation is the secular equation (\ref{secular}).
The function $\zeta_B(k)$ is a complex valued function. It will be convenient
to write it as a real amplitude times a phase factor. Denoting the eigenvalues
of $S_B(k,A)$ by $e^{i\theta _l(k)}$ for $l=1,...,2B$ we get
\begin{equation}
\zeta_B(k)=\exp (\frac i2\Theta (k))2^{2B}\prod_{l=1}^{2B}\sin \frac{\theta
_l(k)}2   \label{tr1}
\end{equation}
where   $\Theta (k)$ is
\begin{eqnarray}
\Theta (k) &\equiv &\frac{1}{i} {\rm log}(\det (-S(k,A)) =
\sum_{l=1}^{2B}\theta _l(k,A)-2B\pi  \nonumber \\
&=&k{\cal L} +(B-V)\pi -2\sum_{i=1}^V \arctan (\Lambda _i) \  .
   \label{tr2}
\end{eqnarray}
 Here ${\cal L}=2\sum_{b=1}^B L_b$ is
twice the total length of the graph, and the $\Lambda_i$ depend on the
boundary conditions as defined in (\ref {lambdadef}). Notice that the
parameters $A_b$ do not appear in the above expression. This is because
the contributions of time reversed bonds are canceled pairwise.

The last product in  (\ref{tr1}) is real on the real $k$ axis. Therefore, the
imaginary part of its logarithmic derivative is a sum of delta distributions
located where $\zeta_B(k) = 0$. Using the expansion
\begin{equation}
\log{\rm det}( I -S_B(k)) = -\sum_{n=1}^{\infty} \frac 1{n}{\rm tr}S_B^n(k)
\label{tr3}
\end{equation}
we obtain the following expression for the density of states
\begin{equation}
d(k)= \frac 1{2\pi }\frac d{dk}\Theta (k)+{\frac 1\pi }{\rm lim}_{\epsilon
\rightarrow 0}\Im m \frac d{dk} \sum_{n=1}^\infty \frac 1n{\rm tr}S_B^n
(k+i\epsilon) \ .  \label{tr3a}
\end{equation}
The first term on the right hand side of   (\ref{tr3a}) corresponds to the
smooth spectral density while the second one provides the fluctuating part.

The spectral counting function $N(k)$ is given by
\begin{equation}
N(k)=\int_0^kd(k^{\prime })dk^{\prime } . \label{tr4}
\end{equation}
 From  (\ref{tr3a}) we have :
\begin{equation}
N(k)=\bar{N}(k)+{\frac 1\pi }\Im m \sum_{n=1}^\infty {\frac 1n}{\rm tr}
(S_B(k))^n
\end{equation}
where
\begin{equation}
\bar{N}(k)-\bar{N}(0)\equiv {\frac 1{2\pi }}\left [\Theta ( k)
-\Theta(0) \right]  ={\frac {k
{\cal L}}{2\pi }} +{\frac V{2}}
-{\frac 1 { \pi }}\sum_{i=1}^V \arctan(\Lambda _i)  \ .  \label{weyl}
\end{equation}
This is the smooth part of the spectral counting function. The leading term
involves the ``volume'' ${\cal L}$ of our system, and it
is independent on the boundary condition $\lambda _i.$ The next two terms
are due to the scattering potentials on the
vertices. The contribution ${\frac V{2}}$ is minus the value of
the third term at $k=0$. For the Neumann boundary condition, the limit
$\lambda_i\rightarrow 0$ should be taken after the value $k=0$ is
substituted.
For large $k$ the last term is inversely proportional to $k$ if $\lambda_i
\ne 0$.
Hence, the mean level density $\bar{d}
=\partial _k\bar{N}(k)$ is essentially constant, reflecting the
fact that the graph is  one dimensional.
 For the Neumann boundary
conditions ($\lambda _i=0$), $\bar{d}$ is independent of the wave number
$k$ and ${\bar{N}(0)}=1/2$ and
\begin{equation}
\bar{N}(k) =  {\frac {k {\cal L}}{2\pi }} + {\frac 1{2}}    \ .
\label{weylHeisenberg}
\end{equation}

The oscillatory part of the counting function is expressed in terms of ${\rm
tr}(S_B(k))^n$. Using the definitions (\ref{DandT}) and $S=DT$ one can obtain
the ${\rm tr}(S_B(k))^n$ directly as sums over $n-$periodic orbits on the
graph:
\begin{equation}
{\rm tr}(S_B(k))^n=\sum_{p\in {\cal P}_n}n_p{\cal A}_p^r{\rm e}^{i(kl_p+
\Phi_p)r}{\rm e}
^{i(\mu _p\pi+\rho_p(k) )r }
\label{posum}
\end{equation}
where the sum is over the set ${\cal P}_n$ of primitive PO's whose
period $n_p$ is a divisor of $n$, with $r=n/n_p$. $l_p = \sum_{b \in p}
 L_{b}$ is the length of the periodic orbit.  $\Phi_p = \sum_{b \in p}L_b
A_b$ is the
 ``magnetic flux" through the orbit. If all the parameters $A_b$ have the same
absolute size $A$ we can write $\Phi_p = A b_p$, where $b_p$ is the directed
length of the orbit.   $\mu _p$ is the number of vertices (with $v_i \geq 2$)
where
 back scattering occurs. At the other $\nu _p$ vertices on the
PO the scattering is not backwards. The number of  back scatters from
vertices with $v_i = 1$
is $n_p-(\mu _p+\nu _p)\ge 0$. The amplitudes ${\cal A}_p$ are given by
\begin{equation}
{\cal A}_p=\prod_{s=1}^{\mu _p}\left| (1-{\frac{2}{v_s(1+i\Lambda_s) }})\right|
\prod_{t=1}^{\nu _p}\left| {\frac{2}{v_t(1+i\Lambda_t) }}\right|\equiv
{\rm e}^{-{\frac{ \gamma _p}2}n_p}  \label{amplitude}
\end{equation}
where $\gamma _p$ plays the r\^{o}le of the Lyapunov exponent.
When $\lambda_i \neq 0$, the phase of each term is modified by adding
\begin{equation}
\label{bcphase}
\rho_p (k) = \sum_{i=1}^{\mu_p}
\arctan\left(\frac{2\Lambda_i}{v_i(1+\Lambda_i^2)-2}\right) -
\sum_{i=1}^{\nu_p}
\frac {\omega_i}{2} - \sum_{i=1}^{n_p-(\mu_p+\nu_p)} \omega_i.
\end{equation}

Substituting (\ref{posum}) in (\ref{tr3a}) one gets an {\it exact} trace
formula
\begin{eqnarray}
d(k) &=&{\frac{{\cal L}}{2\pi }}+{\frac{1}{\pi}}
 \sum_{i=1}^V\frac{v_i\lambda _i} {
(v_ik)^2+\lambda _i^2}    \label{traceformula} \\
&&+{\frac 1\pi }\sum_{p,r}\frac{l_p\cos \left( r(kl_p+\Phi_p+\mu
 _p\pi+\rho_p(k)
)\right) }{{\rm e}^{{\frac{\gamma _p}2}n_pr}}.  \nonumber
\end{eqnarray}

The above formula  bears a striking formal similarity to the well known
exact Selberg trace formula \cite{Selberg} for modular domains on
Riemann surfaces with constant negative curvature, to the Riemann Weyl relation
for the Riemann zeros on the critical line and to the semi-classical Gutzwiller
trace formula for chaotic Hamiltonian systems \cite{G90}. As we shall show
in the sequel, the  analogy between PO's on the graph and  periodic orbits in
dynamical systems follows naturally from the classical dynamics which we
associate
 with the graph. This analogy is  strengthened by further evidence: The
number of $n-$ PO's on the graph is ${\frac 1n}{\rm tr}C^n$, where
$C$ is the connectivity matrix. Since its largest eigenvalue $\Gamma_c$ is
bounded between the minimum and the maximum valency i.e. $\min v_i \leq
\Gamma_c \leq
\max v_i$, periodic orbits proliferate exponentially with topological entropy
$\approx \log \Gamma_c$.  The amplitudes ${\cal A}_p$ which play the
r\^ole of the stability amplitudes,
decrease exponentially with $n$ but not enough to make the series for $d(k)$
absolutely convergent (positive entropy barrier). Finally, $\mu _p$ is the
analogue of the Maslov index. Its origin is topological, and it counts the
number
of non trivial back scatters along the PO. This  can be expressed as the
number of
sequences of strings of the type $\cdots i_a, i_b, i_a\cdots $  (with  $v_b
>2$)
which appear in the code of the PO.

The distinguishing feature of the graph trace formula is  the structure of the
spectrum of lengths $l_p$ of the periodic orbits which appear in
(\ref {traceformula}). In contrast with the other systems mentioned above,
the lengths are constructed as linear combinations (with integer
coefficients) of the elementary bond lengths $L_b$. Hence,
the lengths spectrum is characterized by a high degree of degeneracy.
The degeneracy is not too important for PO's of period $n \le V$.
As $n$ increases, it becomes progressively  dominant, and it is the main
feature
of the length spectrum for periodic orbits with $n>2B$. The effects
of this degeneracy are most apparent in the study of spectral statistics
which will be discussed in the next chapter.


\subsection{{\bf Periodic orbits expression for the spectral }$\zeta$ {\bf
function}}

The spectral density   (\ref{traceformula}) is not   convenient
to deal with, since it is not a proper function but, rather, a distribution.
In this section we would like to cast the information which is stored in
 (\ref{traceformula}) in  a different form, and express it using periodic
orbits.

The first method for achieving this goal is based on the identity (\ref{tr3}),
where the periodic orbit expression (\ref{posum}) for $tr S_B^n$ is
substituted,
and the summation over the repetitions is carried out explicitly.  One gets
\begin{equation}
\zeta_B(k) = \prod_{p} (1-t_p) \ \ ; \ \  t_p= {\rm e}^{-{\gamma_p\over 2}
n_p}
{\rm e}^{i( kl_p+ Ab_p)}{\rm e}^{i(\mu _p\pi+\rho_p(k) ) }.
\label {zeta1}
\end{equation}
This expression as a product over primitive PO's justifies the letter
$\zeta$ by which the secular function is denoted. This product does not
converge on the real $k$ axis because the number of primitive  PO's
proliferates exponentially, with a topological entropy which approximately
equals the mean Lyapunov exponent $\gamma$. Hence, the product (\ref {zeta1})
converges in the absolute sense only for $\Im m\,  k> {\gamma \over 2\left
\langle L\right \rangle }$. This is the ``entropy barrier" for the $\zeta _B$
function. As a matter of fact, the formal manipulations  which were used
to derive (\ref {zeta1}) are strictly justified beyond the entropy barrier.

A converging, well behaved expression for the $\zeta_B$ function on the
real $k$
axis can be obtained, and it is the analogue of the Riemann Siegel
expression for the Riemann $\zeta$ on the critical line.
 To this end, we first note that the $\zeta_B(k)$ function
 (\ref{secular}), is the characteristic polynomial of the matrix $S_B(k)$
\begin{equation}
\label{szf1}
\zeta_B(z,k) = \det (zI-S_B(k)) = \sum_{l=0}^{2B}a_l(k)z^l  \nonumber
\end{equation}
evaluated at $z=1$.
 The coefficients
$a_n(k)$  satisfy an inversive symmetry relation which follows from the
unitarity
of the scattering matrix $S_B(k)$
\begin{equation}
a_{2B-l}(k)=e^{i\Theta (k)}a_l^{*}(k).  \label{ze2}
\end{equation}
Utilizing relation (\ref{ze2}) one may rewrite  (\ref{szf1}) in a more
convenient form
\begin{equation}
\label{szf2}
\zeta_B(z,k)|_{z=1} = 2 e^{i \frac {\Theta (k)}{2}} {\it Re} \left\{ e^{-i
\frac
{\Theta (k)}{2}} \left( \sum_{l=0}^{B-1} a_l + \frac 12 a_B \right) \right\}.
\end{equation}
We would like to emphasize that this form of the secular function
$\zeta_B(k)$ is
due to the unitarity of $S_B(k)$. Thus, the removal of the contributions of
terms with $l>B$ in   (\ref{szf2}) is not only a practical saving of
numerical effort, but also an expression of a basic property of the system.

Let us now consider the $S_B(k)$ matrix at a given wavenumber $k$. Its
spectrum is
on the unit circle i.e $z=\exp(i\theta)$ and $\zeta_B(e^{i\theta},k)$ becomes a
function of $\theta$ which depends parametrically on $k$. Yet, the
secular function (\ref{szf2}) is not real for real $\theta$. It is useful to
define another function which is real on the real $\theta$ line, and
vanishes at $\theta=\theta_l$. Thus, we introduce the spectral determinant
which is obtained from  ~(\ref{szf1}) by extracting a phase factor
\begin{eqnarray}
Z(\theta,k) &= &e^{-i\frac{(\Theta (k)+2B\theta)}2}\zeta_B(e^{i\theta},k) =
2^{2B}\prod_{l=1}^{2B}sin\left(\frac{(\theta_l (k)-\theta)}2\right)\nonumber \\
&=&2  {\it Re} \left\{ e^{-i \frac
{\Theta (k)+2B\theta}{2}} \left( \sum_{l=0}^{B-1} a_l e^{il \theta } +
 \frac 12 a_B e^{i B\theta}\right) \right\}.
\label{szf3}
\end{eqnarray}

The function $Z(\theta,k)$ serves a dual purpose. Setting $\theta=0$ it
is a real secular equation for the graph. For a fixed
value of $k$ it is the secular function for the spectrum of $S_B(k)$.

To write $Z(\theta,k)$ in terms of periodic orbits it is useful
to recall  Newton's identities \cite{S94}
\begin{equation}
a_l=-\frac 1l\left( {\rm tr}S^l+\sum_{n=1}^{l-1}a_n{\rm tr}S^{l-n}\right) ,
\,\,\,\,\,\bigskip\ l=1,...,2B.  \label{ze4}
\end{equation}

The explicit dependence of the $a_l$ on ${\rm tr}S^n$ takes the form \cite
{S94}
\begin{eqnarray}
a_l&=&-\frac 1l {\rm tr}S_B^l  \label{ze5} \\
&-&\frac 1l \sum_{\overrightarrow{l}}^{}\frac{(-1)^n}{ \prod_{i=1}^nl_i}{\rm
tr}S_B^{l-l_1}{\rm tr}S_B^{l-l_1}{\rm tr}S_B^{l_1-l_2}\cdot \cdot \cdot
{\rm tr}
S_B^{l_{n-1}-l_n}{\rm tr}S_B^{l_n}  \nonumber
\end{eqnarray}
where the summation is over all vectors $\overrightarrow{l}$ with integer
entries such that $l>l_1>l_2>\cdot \cdot \cdot >l_n\geq 1.$ By substituting
further from  ~(\ref{posum}) the ${\rm tr}S_B^l$ we get an expression of
the $a_l,\,\,{\rm for}\  l\le B$ in terms of composite orbits \cite{DS92,KB92}.

 The form obtained above for the spectral $\zeta$ function  is reminiscent of
the Riemann-Siegel approximation for the Riemann $\zeta$ on the critical line.
Here, however, it is an exact expression, and because of the fact that the
density of states is constant, the number of terms appearing in the sum is
 independent of $k$.

\subsection{\bf Classical evolution}

We introduced above the concept of orbits on the graph as
 strings of  vertex labels whose ordering obeyed the required connectivity.
This is a finite coding which is governed by a Markovian grammar provided
by the connectivity matrix. In this sense, the symbolic dynamics on the
graph is Bernoulli. So far, the orbits were used and discussed as formal
symbols devoid of a {\it dynamical} origin. In the present section we shall
introduce the classical dynamics which can be associated with a graph, and
which complements the quantum dynamics on the graph. The classical
dynamics makes use of the representation of the graph in terms of directed
bonds. In the following sections we shall use the label $b$ to refer to
the {\it directed} bonds, so that $b=1,\cdots, 2B$ and we shall denote the
time reversed pairs by  $ b =(i,j)\  ;\ \hat b =(j,i)$.

We consider a classical particle which moves freely  as long as it is on a
bond.
The vertices are singular points, and it is not possible to write down
the analogue of the Newton equations at the vertices. Instead, one can employ
a Liouvillian approach, based on the study of the evolution of phase-space
densities. The phase space evolution operator assigns transition probabilities
between phase space points, for which a quantum analogue can be found. The
phase-space description will be constructed on a Poincar\'{e} section which
is defined in the following way. Crossing of the section
is  registered as the particle encounters a vertex, thus the ``coordinate"
on the section is the vertex label. The corresponding ``momentum" is the
direction in which the particle moves when it emerges from the vertex.
This is completely specified by the label of the next vertex to be encountered.
In other words,
\[
\left\{
\begin{array}{c}
{\rm position} \\
{\rm momentum}
\end{array}
\right\} \Longleftrightarrow \left\{
\begin{array}{c}
{\rm vertex}\text{ }{\rm index} \\
{\rm next}\text{ }{\rm index}
\end{array}
\right\} .
\]

The set of all possible vertices and directions is equivalent to the set of
$2B$ directed bonds. The evolution on this Poincar\'{e} section is well
defined once we postulate the transition probabilities $P_{ji\rightarrow
ij^{\prime }}^{(i)}$ between the directed bonds $b=(j,i)$ and $b^{\prime
}=(i,j^{\prime })$. To make the connection with the quantum description,
we adopt the quantum transition probabilities, expressed as the
absolute squares of the $S_B$ matrix elements
\begin{equation}
P_{ji\rightarrow ij^{\prime }}^{(i)}=\left| \sigma_{j,j^{\prime
}}^{(i)}(k,A)\right| ^2=\left| -\delta _{j,j^{\prime }}+{\frac{
(1+e^{-i\omega _i})}{v_i}}\right|^2  \ .  \label{cl1}
\end{equation}
The phases $\omega _i$ are given in~(\ref{smatrix}). For the two
extreme cases, corresponding to Neumann and Dirichlet boundary conditions,
 ~(\ref{cl1}) results
\begin{eqnarray}
P_{ji\rightarrow ij^{\prime }}^{(i)} &=&\left( -\delta _{j,j^{\prime }}+{\
\frac 2{v_i}}\right) ^2,\,\,\,\,\,\,\,\,\,\,\,\text{{\rm Neumann}}  \label{cl2}
 \\
&=&\delta _{j,j^{\prime }}\text{ },\,\,\,\,\,\,\,\,\,\,\,\,\,\,\,\,\,\,\,\,\,\,
\,\,\,\,\,\,\,\,\,\,\,\,\,\,\,\,\,\,\,\,
\bigskip\ \text{{\rm Dirichlet.}}
\nonumber
\end{eqnarray}

The transition probability $P_{ji\rightarrow ij^{\prime }}^{(i)}$ for the
Dirichlet case, admits the following physical interpretation. The particle
is confined to the bond where it started and thus the phase space is divided
to non-overlapping ergodic components ($\approx $``tori''). For all other
boundary conditions, the graph is dynamically connected.

The classical evolution (Frobenius Perron) operator $U_{b,b^{\prime }}$
between the bonds $b,b^{\prime }$ reads
\begin{equation}
U_{ij,nm}=\delta_{j,n} P^{(j)}_{ij\rightarrow jm}
\label{cl3}
\end{equation}
  $U$ does not involve any
metric information on the graph, and for Dirichlet or Neumann boundary
conditions $U$ is independent of $k$.

 If $\rho _b(t)$ denotes the probability to
occupy the bond $b$ at the (topological) time $t$, then we can write down a
Markovian Master equation for the classical density:
\begin{equation}
\rho _b(t+1)=\sum_{b^{\prime }}U_{b,b^{\prime }}\rho _{b^{\prime }}(t).
\label{master}
\end{equation}

The unitarity of the graph scattering matrix $S_B$ guarantees
$\sum_{b=1}^{2B}U_{b,b^{\prime }}=1$ and $0\leq U_{b,b^{\prime }}\leq 1$,
so that the  probability that the particle is on any bond is
conserved during the evolution. The spectrum of $U$ is restricted to the
interior
 of the unit
circle and $\nu_1 = 1$ is always an eigenvalue with the corresponding
eigenvector
$|1\rangle = \frac 1{2B} \left( 1,1,...,1\right) ^T$. In most cases,
the eigenvalue $1$ is the only eigenvalue on the unit circle. Then,
the evolution is ergodic since any initial density will evolve to the
eigenvector $|1\rangle $ which corresponds to a uniform
distribution (equilibrium). The rate at which equilibrium is approached
is determined by the gap between the next largest eigenvalue and $1$. However,
 there are some non generic cases, such as e.g.,
 bipartite graphs when $-1$ belongs to the spectrum. In this case the
asymptotic
distribution is not stationary (see for example Section VII).  If $1$ is
the only
eigenvalue on the unit circle we have
\begin{equation}
\rho (t)\ \overrightarrow{\scriptstyle t{\rightarrow \infty }}\ |1\rangle
\label{epart}
\end{equation}
with a mixing rate $ \ln  \left| \nu _2\right| $ determined by $\nu_2$, the
second largest eigenvalue of $U$. This is characteristic of a classically
mixing system.

Of prime importance in the discussion of the relation between the classical and
the quantum dynamics are the traces $u_n={\rm tr}(U^n)$ which are
interpreted as the mean classical probability to perform $n$- periodic motion.
When only one eigenvalue $\nu$ is on the unit circle, one has that
 $u_n\overrightarrow
{\scriptstyle {n\rightarrow \infty }}\ 1$. Then, we can obtain a classical
sum-rule by substituting the periodic orbit expansion of $u_n$,
\begin{equation}
u_n=\sum_{p\in P_n}n_p\left( |{\cal A}_p|^2\right) ^r\ \ \overrightarrow{
\scriptstyle {n\rightarrow \infty }}\ 1.  \label{classicalsum}
\end{equation}
Each periodic orbit is endowed with a weight $|{\cal A}_p|^2$ defined in
terms of
the stability amplitudes (\ref{amplitude}). It is the probability to remain
on the orbit. These weights are the counterparts of the stability weights $%
|\det (I-M_p)|^{-1}$ for hyperbolic periodic orbits in Hamiltonian systems,
where $M_p$ is the monodromy matrix. Graphs, however, are one dimensional
and the motion on the bonds is simple and stable. Ergodic (mixing) dynamics
is generated because at each vertex a (Markovian) choice of one out of $v$
directions is made. Thus, {\it  chaos on graphs originates from the
multiple connectivity of the (otherwise linear) system.}

Using the expression (\ref{classicalsum}) for $u_n$ one can easily write
down the complete thermodynamic formalism for the graph. Here, we shall only
quote the periodic orbit expression for the Ruelle $\zeta $ function
\begin{eqnarray}
\zeta _R(z) &\equiv &\left( \det (I-zU)\right) ^{-1}={\rm \exp }\left[ -{\rm
tr}\left( \ln (I-zU)\right) \right]  \label{cl4} \\
&=&\exp \left[ \sum_n\frac{z^n}nu_n\right] =\prod_p\frac 1{\left(
1-z^{n_p}\exp (-n_p\gamma _p)\right) ^{}}  \nonumber
\end{eqnarray}
where the product extends over all primitive periodic orbits and we have
used the definitions of   (\ref{amplitude}).


\section{\bf Spectral Statistics}

So far we developed the spectral theory of graphs, pointing out the similarity
between quantum graphs and more complex quantum systems which display chaotic
dynamics in the classical limit. In the present chapter we shall report about
analytical and numerical results which show that the spectral statistics of
these
simple systems also follow the pattern of more general Hamiltonian systems.
Namely, when the classical graph dynamics is mixing, and in the limit when the
(topological) time needed to reach equilibrium is much shorter than  the number
of directed bonds, the spectral statistics for quantum graphs are very well
reproduced by the predictions of Random Matrix Theory. In the integrable limit
(Dirichlet boundary conditions) the graph spectral statistics is Poisson as
is the case in generic integrable Hamiltonian systems. The investigation of
the universality of spectral fluctuations and deviations thereof is especially
convenient for graphs because of the transparent and simple spectral theory
in terms of PO's, and because of the relative numerical ease by which large
spectral data bases can be constructed. The parameters which appear in the
theory can be used to study various characteristic spectral transitions:
Changing the vertex potential parameters $\lambda_i$ (which, for simplicity, we
will choose to be the same for all the vertices i.e.$\lambda=\lambda_i  $)
induces the transition between classical integrability to chaos. This is
accompanied by a spectral transition from Poisson to RMT like statistics.
The parameter $A_b$ (again for the sake of clarity we will choose $|A_b|=
A $ for all the bonds) allows to break time reversal symmetry, and the
mean valency can be used to study the dependence of spectral statistics on
connectivity. All these questions will be dealt with in the present chapter.

The spectral theory presented above relied heavily on the bond scattering
matrix $S_B$ which provided the foundation for the periodic orbits theory
and the connection to the classical evolution. The spectral statistics of
the eigenphases of the $S_B$ matrix  are also intimately connected with the
spectral statistics of the graph wavenumber spectrum. Therefore, we will
start our presentation by discussing the $S_B$ matrix spectral statistics.

\subsection{{\bf The spectral statistics of the $S_B$ matrix}}

We consider the $S_B(k,A)$ matrices defined in  (\ref {secular}, \ref{scaco}).
Their spectrum consist of $2B$ points confined to the unit circle
(eigenphases).
Unitary matrices of the type considered here are commonly studied since they
are the quantum analogues of classical, area preserving maps. Their spectral
fluctuations depend on the nature of the underlying classical dynamics
 \cite{S89}.
The quantum analogue of classically integrable maps display Poissonian
 statistics
while in the opposite case of classically chaotic maps, the eigenphases
 statistics
conform quite accurately with the results of RMT for Dyson's {\it circular
 ensembles}.
The ensemble of unitary matrices which will be used for the statistical study
will be    the set of matrices $S_B(k,A)$ with $k$ in the range $|k-k_0|
\le \Delta_k/2$, where the mid point $k_0$ and the interval size $\Delta_k$
are to be determined. Since the dimension of the $S_B$ matrices is independent
of $k$, the mean value $k_0$ is important only when the boundary conditions
are neither Neumann nor Dirichlet. For the intermediate boundary conditions,
$k_0$ sets the mean value of the parameter  $\Lambda $ (\ref {lambdadef}),
 and $\Delta_k$ must be
chosen such that   $\Lambda  $   does not change appreciably in the interval.
However, $\Delta_k$ must be much larger than the correlation length between
the matrices  $S_B(k,A)$. One can estimate  the correlation length by
studying  the auto-correlation function
\begin{equation}
C(\chi )\equiv\frac 1{\Delta_k}\int_{k_0-\Delta_k/2}^{k_0+\Delta_k/2}
 \frac 1{2B}{\rm tr}\left( S_B^{\dagger }(k^{\prime }+\frac \chi
2)S_B(k^{\prime }-\frac \chi
2)\right) dk^{\prime }.  \label{enscor1}
\end{equation}

For the two extreme cases of Neumann and Dirichlet boundary conditions the
auto-correlation function (\ref{enscor1}) can be calculated exactly. By writing
the scattering matrix as $S_B(k)=exp(ikL)S_B(0)$ and substituting in
(\ref{enscor1})
 we find:
\begin{equation}
C(\chi )=\frac 1B\sum_{i=1}^Be^{iL_b\chi }=\int e^{iL\chi }P(L)dL
\label{enscor2} \\
\equiv {\hat{P}}(\chi)
\end{equation}
where ${\hat{P}}(\chi)$ is the Fourier transform of
the probability distribution $P(L)$ of the lengths of the bonds. Thus, the
correlation is inversely proportional to the variance of the distribution of
the lengths $L_b$. From now on, we  shall  assume
 $\Delta_k \gg {\rm var} (L_B)$, which justifies the $k$ averaging
 procedure.  The ensemble average will be denoted by
\begin{equation}
\label{ensaveS}
\left \langle \ \cdot \ \right \rangle _k \equiv \frac {1}{\Delta_k}
 \int_{k_0-\Delta_k/2}^{k_0+\Delta_k/2}\cdot
\,\, dk \ .
\end{equation}

 Another  way to generate an ensemble of graphs, is by
randomizing the length matrix $L$ which contains the lengths of the bonds
while the connectivity (topology of the graph) is kept constant. This is the
disorder approach, which will also be applied when called for.

In the following subsections VI.A.1-2, we investigate some statistical
measures
 of the eigenphases $\left\{ \theta _l(k)\right\} $ of
the scattering matrix $S_B$ \cite{M90,BS88} and compare them  with the
predictions of RMT, and with the results of
the periodic orbits theory of spectral fluctuations
which was originally developed for quantized maps. The two
statistics which we shall investigate are the spectral form factor
and  the autocorrelation of the
spectral $\zeta$ function. Explicit expressions for these quantities are given
by Random Matrix Theory \cite{HKSSZ96} and a  semiclassical  theory is also
available \cite{keatbog,UScorr,camb}.


\subsubsection{\bf The Form Factor}

The $S_B$ matrix for a fixed eigenvalue $k$ is a unitary matrix with
 eigenvalues $e^{i\theta_l(k)}$.  The
 spectral density of the eigenphases then reads
\begin{equation}
d(\theta;k )\equiv \sum_{l=1}^{2B}\delta (\theta -\theta
_l(k))=\frac{2B}{2\pi }+
\frac 1{2\pi }\sum_{n=1}^\infty e^{-i\theta n}{\rm tr}S_B(k)^n+{\rm c.c}
\label{sms1}
\end{equation}
where the first term on the r.h.s is the smooth density $
\overline{d}=\frac{2B}{2\pi }$, while the others describe the fluctuating
part.

The two-point correlations are expressed in terms of the excess probability
density
$R_2(r)$ of finding two phases at a distance $r$, where $r$ is measured in
units of the mean spacing  ${ 2\pi \over 2B}$
\begin{equation}
R_2(r;k_0)={2\over 2\pi}\sum_{n=1}^\infty \cos \left(
\frac{2\pi rn}{2B}\right) \frac 1{2B}\left<\left| {\rm
tr}S_B^n\right|^2\right>_k\,\, .
\label{sms3}
\end{equation}

The form factor $K(n,2B)={\frac 1{2B}}<|{\rm tr}S_B^n|^2>_k$ is the Fourier
transform of $R_2(r,k_0)$. For a Poisson spectrum, $K(n,2B)=1$ for all $n$.
 RMT predicts that  $K(n,2B)$, depends on the
scaled  time $\frac {n}{2B}$ only \cite{S89}, and explicit expressions
for the orthogonal and the unitary circular ensembles are known \cite{HKSSZ96}.

We computed ${\frac 1{2B}}<|{\rm tr}S_B^n|^2>_k$  for well connected graphs,
with various vertex potential parameters. In Fig.~1 we show  typical examples,
calculated for a fully connected pentagon. The results for Neumann boundary
conditions  show  quite a good  agreement  with the predictions of RMT for
the Circular ensembles. We shall discuss and explain these results in the
following paragraphs.

To begin, consider  the data for Neumann boundary conditions and $A=0$ or
$A\ne 0$ (see Fig.~1). The predictions of RMT are also shown, and they
reproduce quite well the smooth trend of the data in the two cases. The
deviations from the smooth curves are not statistical, and cannot be ironed
out by further averaging. Rather, they are due to the fact that the graph is a
dynamical system which cannot be described by RMT in all detail. To study
this point in depth we shall express the form factor in terms of the PO
expression ~(\ref{posum}). Assuming Neumann boundary conditions for the
time being,
\begin{eqnarray}
 \frac 1{2B}\left\langle \left| {\rm tr}S_B^n(k)\right| ^2\right\rangle_k &=&
\frac 1{2B}\left\langle \left|\sum_{p\in {\cal P}_n}n_p{\cal A}_p^re^{i(kl_p+A
b_p+ \pi \mu
_p)r}\right|^2 \right\rangle_k  \label{sms5} \\
&=&\left . \frac 1{2B} \sum_{p,p'\in {\cal P}_n}  n_pn_{p\prime} {\cal A}_p^r
{\cal A}_{p\prime}^{r^{\prime}}
\exp \left  \{ iA(r b_p-r'b_{p\prime}) +i\pi (r\mu_p-r'\mu_{p\prime})\right\}
 \right |_{rl_p = r^{\prime}l_{p^{\prime}}} \nonumber \\
\nonumber
\end{eqnarray}
The $k$ averaging is carried out on such a large interval that the double
sum above is restricted to pairs of periodic orbits which have exactly the
same length. The fact that we choose the lengths of the bonds to be rationally
independent will enter the considerations which follow in a crucial way.
Consider first the domain $1 <n \ll 2B$. The PO's are mostly of the
irreducible type, and the length restriction limits the sum to pairs of
orbits which are conjugated under time reversal. Neglecting the contributions
from repetitions and from self tracing orbits we get
\begin{equation}
\label{transuzy1}
\frac 1{2B}\left\langle \left| {\rm tr}S_B^n(k)\right| ^2\right\rangle_k
\approx \frac 1{2B} \sum_{p\in {\cal P}_n}  n^2 {\cal A}_p^2 \ \
4\cos ^2A b_p  = {2n\over 2B} u_n \left \langle \cos ^2  A b_p\right
\rangle _n \ .
\end{equation}
The classical return probability $u_n$ approaches $1$ as $n$ increases
(see (\ref{classicalsum})). However, deviations from unity reflect
the fact that the classical dynamics reaches the ergodic state only after
some time. The  deviation which is simplest to understand occurs at $n=1$.
Since there are no classical fixed points (no self connected vertices) on
the graph, $u_1=0$. However in the limit $B\rightarrow \infty$, the short
time deviations converge to the origin when the scaled form factor is studied.
Neglecting the short time deviations, we can replace $u_n$ by $1$, and we
see that the remaining expression to be evaluated is the classical expectation
of $\cos ^2  A b_p$ over PO's of length $n$. For $A=0$ this factor
is identically $1$ and one obtains the leading term of the COE expression
for $n\ll 2B$. If $A$ is sufficiently large $ \left \langle \cos ^2  A b_p
\right \rangle_n \approx 1/2 $, and one obtains the short time limit of the
CUE expression. The transition between the two extreme situations is well
described by
\begin{equation}
\label{transuzy2}
 \left \langle \cos ^2  A b_p\right \rangle _n \approx {1\over 2} \left (
 e^ {- A^2\left \langle L_b^2\right \rangle {n\over 2}} +1 \right ) \ .
\end{equation}
This formula is derived by assuming that the total directed length $b_p$ of a
periodic orbit is a sum of elementary lengths with random signs.

One cannot use the arguments presented above for the range  $ n \ge B$. As $n$
approaches $B$ the degeneracy of the length spectrum prevails and for
$n>2B$ all the orbits are degenerate. In other words, the restriction
$rl_p = r^{\prime}l_{p^{\prime}}$ in (\ref {sms5}) does not pick up a unique
orbit and its time reversed partner, but rather, a group of isometric but
distinct orbits. Therefore, the interferences of the contributions from the
group of all the orbits must be calculated. The relative sign of the terms
is determined by the ``Maslov" index. This can be seen better, once rewriting
(\ref{sms5}) in the form (we assume for simplicity $A=0$)
\begin{eqnarray}
\frac 1{2B}\left\langle \left| {\rm tr}S_B^n(k)\right| ^2\right\rangle_k
& = &
\frac 1{2B}\sum_{\{q\}} \left|\sum_{p\in {\cal P}_q}n_p{\cal A}_p^re^{i\pi
 \mu_pr}
\right|^2, \,\,\,\,\,\,
\label{masnew}
\end{eqnarray}
where the second sum is over the set ${\cal P}_q$ of PO of the type $l_p =
\sum q_b L_b$ with $\sum q_b = n_p$. It is clear that the indices of different
orbits in a family of isometric PO's are {\it correlated}. Otherwise, if the
Maslov indices are random, one would regain the diagonal approximation
(\ref {transuzy1}) for  arbitrarily long times. The correlation  between the
Maslov indices within the family of isometric PO's are the analogue  of action
correlations in the semiclassical theory of spectral statistics \cite
 {ADDKKSS93}.
The dynamical origin of these correlations is not known also for graphs, and it
is one of the important open problems that should be addressed.

 Graphs with Dirichlet boundary conditions are integrable in the sense
explained above. One expects therefore that the spectral statistics in
this case is Poissonian, which implies
\begin{equation}
\frac 1{2B}\left\langle \left| {\rm tr} S_B^n(k)\right| ^2\right\rangle_k = 1
\ \ \
{\rm for \ \ all} \ \ \ n>1 \ .
\end{equation}
 In the Dirichlet case, the $S_B$ matrix reduces to a block diagonal form where
each bond and its time reversed partner are coupled by a $2\times 2$ matrix
of the form
\begin{eqnarray}
S^{(b)}(k,A) =\left (
{\begin{array} {ll}
     0  &   e^{i(k+A)L_b}  \\
    e^{i(k-A)L_b }& 0
\end{array}}
\right ) \  .
\end{eqnarray}
The spectrum of each block is the pair  $\pm e^{ikL_b}$, with the
corresponding
symmetric and antisymmetric eigenvectors $ {1\over \sqrt {2}}(1, \pm1)$.
As a result, we get
\begin{equation}
\frac 1{2B}\left\langle \left| {\rm tr} S_B^n(k)\right| ^2\right\rangle_k
 =1+(-1)^n \ \ \
{\rm for \ \ all} \ \ \ n\geq 1 \ .
\label{poissonform}
\end{equation}

 This deviation from the expected Poissonian result is due to the fact that
the extra symmetry reduces the $S_B$ matrix further into the symmetric
and antisymmetric subspaces. The spectrum in each is Poissonian, but when
combined together, the fact that the eigenvalues in the two spectra differ
only by a sign is the reason for the anomaly (\ref {poissonform}). In the
sequel we shall remove this feature by considering the smooth form factor
obtained by taking the mean of successive $n$ values.

 The transition between the two extreme boundary conditions can be affected
by using the interpolating boundary conditions where $\lambda \neq \{ 0 ,
\infty\}$.
The relevant parameters are the $\Lambda$ defined in (\ref {lambdadef}),
and it is expected that the spectral statistics make the transition from
RMT like to Poisson as these parameters spans their range of values. The exact
symmetry which prevails in the Dirichlet case, is broken for intermediate
values of $\Lambda $. However, the tendency towards trapping in a single
bond is a dynamical feature, which persists for finite $\Lambda$ values,
and therefore the even-odd staggering of the form factor can be observed also
for the intermediate range of $\Lambda$ values (see inset in Fig.~1).
Since the dynamical reason for this effect is clearly understood, we show in
Fig.~1 the pairwise averaged form factor, which displays clearly the
transition from the Poissonian to the RMT (COE and CUE) limit.

 As was mentioned above, the short times deviations of the data from the
RMT expectations (see Fig.~1) are real, and are due to the deterministic nature
of the dynamics induced by the $S_B$ matrices. It is easy to show this
explicitly for  $\left<|{\rm tr} S^2|^2\right>$, since here  all the
contributions
are due to period-$2$ PO's which are self tracing, and each has its
distinct length.
Using  (\ref {posum}) we get
\begin{equation}
\label{tsampi}
\frac 1{2B}\left\langle \left| {\rm tr} S_B^2(k)\right| ^2
\right\rangle_k = 2 \left (\frac{(1-\frac 2v)^2 +\Lambda^2}{1+\Lambda^2}
\right )^2 \ ,
\end{equation}
  independently of the value of $A$. This is different from the value $
{1\over B}$
 expected for the CUE and $\approx {2\over B}$ expected for the COE.

\subsubsection{\bf Spectral $Z$ function correlations}

We conclude this section with an analysis of the spectral $Z$ function
given by ~(\ref{szf3}). The statistical properties of this function, can
be expressed in terms of the statistics of either the eigenphases $\theta_l$
or the coefficients $a_{l}$ (see (\ref{szf3})). Since the two sets of variables
are functionally related, they are statistically equivalent. In practice,
however, one cannot check the full spectral distribution, and therefore it is
advantageous to study statistical measures which are based on other accessible
quantities. The measure which was proposed in \cite{UScorr} and \cite{KKS97}
was the autocorrelation function
\begin{eqnarray}
C_{Z}(\eta ) &\equiv&\int_0^{2\pi }\left\langle Z(\theta+\frac \eta
2;k)
Z^{\star} (\theta- \frac \eta 2;k)\right\rangle_k \frac {d\theta}{2\pi}
\label{sms6}
 \\
&=&\sum_{l=0}^{2B}\left\langle \left| a_l\right| ^2\right\rangle_k e^{i\eta
(l-B)}.  \nonumber
\end{eqnarray}
This statistical measure, depends on higher order correlations of the
eigen-phases. Hence, $C(\eta )$ tests aspects of the eigenvalues distribution
which are not accessible by the study of the two-point form factor discussed
previously.

The ensemble averages $\left\langle \left| a_l\right| ^2\right\rangle $ for
Circular Random Matrices were calculated in \cite{HKSSZ96}. We shall quote
here the results for the COE and CUE ensembles:
\begin{eqnarray}
\left\langle \left| a_l\right| ^2\right\rangle _\beta \ \  &=&1+\frac{l(2B-l)}{
2B+1}\ ,\,\,\bigskip\ \beta =1  \label{sms7} \\
&=&1\ ,\,\,\,\,\,\,\,\,\,\,\,\,\,\,\,\,\,\,\,\,\,\,\,\,\,\,\,\,\,\,\,\,\,\,
\bigskip\ \beta =2  \nonumber \\
&&  \nonumber
\end{eqnarray}

An approximate expression for the $\left\langle \left| a_l\right|
 ^2\right\rangle$
was obtained by assuming that the  ${\rm tr}S^n$ are independent random
Gaussian variables for $n<B$ \cite{UScorr}. This approximation
is an extension of the diagonal approximation mentioned above, and
it leads to the following  recursion relation
\begin{equation}
\label{rr}
\left\langle \left| a_l\right| ^2\right\rangle = \frac 1{l} \sum_{n=1}^{l}
\left\langle \left| a_{l-n}\right| ^2\right\rangle \frac
{\left\langle \left| {\rm tr}S_B^n\right| ^2\right\rangle}{n}.
\end{equation}
For the calculation of the autocorrelation function (\ref{sms6}), it
is sufficient to obtain the $\left\langle \left| a_l\right| ^2\right\rangle$
for $l\leq B$. The rest are provided by the inverse symmetry (\ref{ze2}) which
moreover implies that the Fourier components of $C(\eta )$ are symmetric about
$B$ and thus the autocorrelation function is real.

Using the approximate result (\ref{transuzy1}) for the form factor
$\left\langle \left| {\rm tr}S_B^n\right| ^2\right\rangle \simeq gn u_n $
we have \cite{KKS97}
\begin{equation}
\left\langle \left| a_l\right| ^2\right\rangle =\frac gl\sum_{k=1}^l\left
\langle \left| a_{l-k}\right| ^2\right\rangle u_{k,}  \label{sms8}
\end{equation}
which should be solved with the initial condition $\left\langle \left|
a_0\right| ^2\right\rangle =1$. In (\ref{sms8}) $g=2(1)$ for systems with
(without) time reversal symmetry, and $u_k= {\rm tr}U^k$ is the classical
return probability. For systems which display strong mixing,
$u_n =1$, and the approximate recursion relations reproduce the RMT
result for systems which violate time reversal symmetry ($\beta =2$).
For systems which are invariant under time reversal, one reproduces
only the leading term in ${n \over 2B}$ of the RMT result ($\beta=1$).

We computed numerically the $\left \langle|a_l|^2\right \rangle _k$ for
the completely connected  pentagon, subject to Neumann boundary conditions,
where  time reversal symmetry was either respected ($A=0$) (Fig.~2a), or
 violated
($A\ne 0$) (Fig.~2b). The results are displayed in Fig.~2, and they deviate
substantially from the RMT predictions (\ref{sms7}). Note that  the values of
$\left\langle \left|{\rm tr}S_B^n\right|^2 \right \rangle_k$ for the same
system showed a rather good agreement with RMT (see Fig.~1). The reason
for the large deviation is clear. No physical system can reproduce
the strong mixing condition $u_n=1$ for all $n$. Indeed, this is the reason
why $\left\langle \left|{\rm tr}S_B^n\right|^2 \right \rangle_k$ deviate
from the RMT results for short times.
Because of the iterative procedure (\ref{sms8}), the short
time non-generic effects reveal  themselves in the higher order coefficients
$a_l$, and this is why this statistical measure is much more sensitive
to the non universal features of the classical dynamics. The approximate
theory presented above, includes the correct short time behavior
of the system, and therefore it reproduces the main
features of the numerical data much better than RMT. A
quantitative measure for the expected deviation from the RMT prediction
can be given by the magnitude of the next to the leading eigenvalue of the
 classical evolution operator $U$. For the
system we studied, it  is -.25, which is still far from the value 0 expected
in the strong mixing  limit.

Let us finally comment on the transition from Poisson to RMT due to variation
of the parameters $\Lambda_i$. As it was mentioned already in the introduction,
for integrable systems, we expect that the spectrum is uncorrelated and
 described
by the Poisson ensemble which gives for the $\langle \left| a_{l}\right|^2
\rangle $
the expression
\begin{equation}
\label{apois}
\langle \left| a_{l}\right| ^2\rangle =
\left (
\begin{array}{c}
2B\\l
\end{array}
\right ).
\end{equation}
On the other hand, the diagonal approximation predicts that
$\left \langle|{\rm tr} S^n|^2 \right \rangle \simeq 2B$
(see previous section) and thus it provides us with the following recursion
 relation
for the coefficients $\langle \left| a_{l}\right| ^2\rangle$
\begin{equation}
\label{apoissem}
\langle \left| a_{l}\right| ^2\rangle = \frac{2B}{l}\sum_{n=1}^l
\frac{\langle \left| a_{l-n}\right| ^2\rangle}{n}.
\end{equation}
In Fig.~2a,b we present our numerical results for a fully connected pentagon
and for various values of the parameter $\Lambda$. Again, we see that our
system
undergoes a transition from GOE/GUE to Poisson statistics when $\Lambda$
 increases.


\subsection{\bf Level Statistics}

The statistical properties of the energy levels (or the wave numbers) spectrum
can be derived from  the statistics of the eigenphases of $S_B$ because of the
following reasoning. The wavenumber spectral density can be written as
\begin{equation}
d(k) = \sum_{n=1}^{\infty} \delta (k-k_n) = \sum_{l=1}^{2B}\delta_{2\pi}
\left (\theta_l(k)\right ) \left |{{\rm d}\theta_l(k)\over{\rm d}k} \right |
\label {d(k)}
\end{equation}
One can easily show that
\begin{equation}
   L_{min}\le {{\rm d}\theta_l(k)\over{\rm d}k} \le L_{max}
\label{lbound}
\end{equation}
 where $L_{min,max}$ denote the minimal or the maximal bond length
of the graph. Consider an interval $\delta k$ about $k_0$ so that
$\left \langle {{\rm d}\theta_l(k)\over{\rm d}k}\right \rangle _l \delta k =
\left \langle L \right \rangle   \delta k < 2\pi $. Since the mean wavenumber
 spectral density is
${\left \langle L \right \rangle   B \over \pi} $ the interval $\delta k$ can
 accommodate a large
number of levels when $B$ is large. The wavenumber density in the $\delta k$
vicinity of $k_0$ is
\begin{equation}
d(k;k_0) = \sum_{l=1}^{2B}\delta_{2\pi}
\left (\theta_l(k_0) +(k-k_0){{\rm d}\theta_l(k)\over{\rm d}k} \right )
  {{\rm d}\theta_l(k)\over{\rm d}k}
\approx \left \langle L \right \rangle \sum_{l=1}^{2B}\delta_{2\pi}
\left ( \theta_l(k_0) +(k-k_0) \left \langle L \right \rangle  \right )
\end{equation}
For a given $k$ value, the expression on the rhs is the eigenphase density
of the unitary matrix $S_B(k_0)$, (the $l$ independent shift of the phases
does not change the distribution of intervals on the circle). Hence, one can
read the short range statistical properties of the $k$ spectrum, from
the results on the statistics of the eigenphases which was discussed in
the previous section. In the sequel we shall supply numerical
data and additional arguments to show that this is indeed the case. We shall
also compute various statistical measures which are commonly used in the
statistical  analysis of spectral fluctuations of quantum systems whose
classical analogue is chaotic. We shall show that the spectrum of the
quantized graph behaves as a typical member of this set of ``quantum-chaotic"
systems.

For the numerical calculation of the spectrum we had used the method described
in section III.A. That is, we identified the spectrum as the zeros of $\det
 h(k,A)$.
The  completeness of the spectrum was checked by comparing the counting
function
$N(k)$ with Weyl's law ~(\ref{weyl}). An efficient detector of missing or
 spurious
levels is provided by the function $\delta _n$
\begin{equation}
\delta _n=N(k)-\overline{N}(k) \ .  \label{nume4}
\end{equation}
This quantity is expected to fluctuate around zero and a  redundant or a
missing eigenvalue is accompanied by an offset by $\pm 1$.
Fig.~3 shows a typical plot of $\delta _n$ as a function of $n$ for
a graph containing 5 vertices (see inset in Fig. 3) and Neumann boundary
conditions. $\delta_n$ fluctuates about $0$ as expected, with $|\delta_n|<2$
which is a quantitative indication of the rigidity of the spectrum.
This is the behavior expected for a quantum chaotic system.

The spectral fluctuations are best studied in terms of the
 unfolded spectrum $\left\{ x_n\right\} $
\begin{equation}
x_n=\overline{N}(k_n)  \label{unfold}
\end{equation}
whose mean level spacing is unity. Since (\ref{weyl}) provides an exact
expression for $\overline{N}(k)$ the unfolding procedure is straight
forward.


\subsubsection{\bf Level spacing distribution}

The distribution $P(s)$ of the spacings $s_n=x_{n+1}
-x_n$ of adjacent quantal levels (or its integrated form
$I(s) = \int_0^{s} P(r) {\rm d}r$)
is  the most convenient and commonly used statistics. The expression
for $P(s)$ for the Poisson, GOE and GUE ensembles are well known, and have been
compared with the distributions derived from the graph spectra. The numerical
results for many systems show that the graphs follow the general trend
observed for realistic systems. As a typical example, the results for the
fully connected quadrangle with Neumann boundary conditions ($\lambda
_i=0)$ and
with $A=0$ and $A\neq 0$ are shown in Fig.~4. They are based on the leading
80,000 eigenvalues for each case. We would like to emphasize that the spectra
were calculated for a fixed set of bond lengths, in other words, no disorder
averaging was employed. The agreement with the exact (not the Wigner surmise)
 RMT
curves is very good \cite{DH90}, although systematic deviations at the level
of $1\%$ or less can be discerned (see inset of Fig.~4).
These differences exceed the statistical error margin, and we believe that
they originate from the fact that the short time dynamics on the graph does
not follow the universal pattern, as  was explained in the previous section.

We have already noted that the Poisson limit is obtained naturally for
graphs which are subject to Dirichlet boundary conditions. The transition
between the two extremes is affected by changing the parameter $\Lambda =
{\lambda_i\over v_i k  }$.  Using this time a fully connected pentagon, we
observe the transition in  the nearest neighbor distribution as is shown in
Fig.~5a,b for $A=0$ and $A\neq 0 $ respectively.

We made similar comparisons for other well connected graphs and observed the
same degree of agreement between the data and the results of RMT. Thus, we
face an exceedingly simple class of systems which, according to
the nearest neighbor statistics,  belongs to the same spectral
universality class as quantum systems which are chaotic in the classical
limit. We shall study below other statistical measures, and show that
deviations from universality appear as expected and observed in generic
Hamiltonian systems.


\subsubsection{\bf The form factor}

To investigate further the dynamical origins of the level fluctuations we
study the two point form factor $K(\tau ;k_0)$
 defined by
\begin{equation}
K(\tau ;k_0)={\frac 1{{\cal N}}}\left| \sum_{|k_n-k_0|\le \Delta
_k/2}{\rm e}^{i\ 2\pi k_n{\cal L}\tau }\right| ^2-{\cal N}\delta (\tau )
\label{k(t)}
\end{equation}
where we consider a spectral interval of size $\Delta _k$, centered about $
k_0$ and involving ${\cal N}=\bar{d}\Delta _k$ eigenvalues. $\tau $
measures lengths in units of the Heisenberg length $l_H={\cal L}$. The main
reason of our choice to base our analysis on the two point form factor is
that it allows us a study of the level fluctuations in terms of PO's.
Indeed, recalling that
\begin{equation}
\sum_{|k_n-k|<\Delta_k/2} exp(2\pi i k {\cal L} \tau) \equiv \int_{k_0 -
\Delta_k/2}^{k_0 +\Delta_k/2} d(k)exp(2\pi i k {\cal L} \tau) dk
\label{2spco}
\end{equation}
and expressing $d(k)$ by its PO expansion given by ~(\ref
{traceformula}) we can rewrite $K(\tau ;k_0)$, after substituting the
resulting expression into  ~(\ref{k(t)}), in terms of periodic orbits.

 We shall concentrate for the time being on graphs with
Neumann boundary conditions.  Splitting $K(\tau ;k_0)$
to its diagonal $K_D(\tau;k_0)$ and non-diagonal parts $K_{ND}(\tau ;
k_0)$, we write them in terms of periodic orbits and their repetitions
\begin{eqnarray}
K_D(\tau ;k_0) &=&{\frac{2{\cal N}}{{\cal L}^2}}\sum_{p;r}|\tilde{\cal A}%
_p^r|^2\ \ \left( \delta _{{\cal N}}({\frac{rl_p}{{\cal L}}}-\tau )\right) ^2
\label{k(t)po} \\
K_{ND}(\tau ;k_0) &=&{\frac{2{\cal N}}{{\cal L}^2}}\sum_{p,r\ne
p^{\prime },r^{\prime }}\tilde{\cal A}_p^r\ \tilde{\cal A}_{p^{\prime
}}^{r^{\prime
}}\ {\rm e}^{i\pi (r\mu _p-r^{\prime }\mu _{p^{\prime }})}{\rm e}%
^{iA(rb_p-r^{\prime }b_{p^{\prime }})}  \nonumber \\
&\times &\ \delta _{{\cal N}}({\frac{rl_p}{{\cal L}}}-\tau )\ \delta _{{\cal %
N}}({\ \frac{r^{\prime }l_{p^{\prime }}}{{\cal L}}}-\tau )\cos k_0%
(rl_p-r^{\prime }l_{p^{\prime }})  \nonumber
\end{eqnarray}
where we use $  \tilde{\cal A}_p=n_pl_p{\cal A}_p$ and $\delta _{{\cal
N}}(x)={\ \frac{%
\sin {\frac{{\cal N}x}2}}{{\frac{{\cal N}x}2}}}$. $K_D$ is a classical
expression, because all interference effects are neglected, but for the ones
which are due to exact symmetries. The sum-rule (\ref{classicalsum}),
enables us to justify a Hannay and Ozorio de Almeida -like sum rule \cite
{HOA}, namely,  $K_D(\tau )\approx \langle g\rangle \tau $
\cite{berry}. ($\langle g\rangle$ is the mean degeneracy of the length
spectrum due to exact symmetries such as time reversal).
For $\tau \ll 1$ $K(\tau )\approx K_D(\tau )$.
Because of the fact that the quantum spectrum is real and discrete,
$K(\tau )$ must approach $1$ for $\tau >1$. This is taken care of by
$K_{ND}$. In contrast to the diagonal part, $K_{ND}$, depends crucially on
the phase correlations between the contributing terms. Actually, its
Fourier transform tests how the $l_p$ spectrum is correlated
\cite{ADDKKSS93}. In Hamiltonian systems in more than one dimension, the
size of the spectral interval $\Delta _k$ is limited by the requirement that
the smooth spectral density is approximately constant. Here $\bar{d}$ is
constant, hence one can take arbitrarily large $\Delta _k$. This way, one
can reach the domain where the function $K(\tau )$ is composed of
arbitrarily sharp spikes ($\delta _{{\cal N}}(x)$ can become arbitrarily
narrow) which resolve completely the length spectrum for lengths which are
both smaller and larger than $L_H$. In Fig.~6a (Fig.~6b) we show the
numerical $K(\tau )$ calculated with two extreme values of ${\cal N}$ for
the case with (without) time reversal invariant symmetry. As long as $\tau
{\cal L}$ is shorter than the length of the shortest period orbit, $K(\tau
)=0$, while for $\tau >1$ it saturates and fluctuates around the value one.
The RMT two-point form factor given as \cite{M90}
\begin{eqnarray}
K_{GOE}(\tau ) &=&\left\{
\begin{array}{cc}
2\tau -\tau \ln (1+2\tau ), & 0\leq \tau \leq 1\nonumber \\
2-\tau \ln \left( \frac{2\tau +1}{2\tau -1}\right) , & \tau \geq 1
\end{array}
\right\}  \\
K_{GUE}(\tau ) &=&\left\{
\begin{array}{cc}
\tau , & 0\leq \tau \leq 1 \\
1, & \tau \geq 1
\end{array}
\right\}   \label{rmtff}
\end{eqnarray}
is also shown in Figs.~6a and 6b for comparison. Despite the fluctuations,
the low resolution curve does not deviate much from the prediction of RMT.
The high resolution data shows a similar behavior, which can be better
checked if one studies the integrated form factor (see inset of Figs.~6a,b)
\begin{equation}
\tilde{K}(\tau )={\frac 1\tau }\int_0^\tau K(t){\rm d}t.  \label{iff}
\end{equation}
However, by increasing the resolution, correlations between periodic orbits
with different lengths are suppressed, and the interference mechanism which
builds up $K_{ND}$ cannot be due to the correlations in the spectrum of
periodic orbit lengths, but to another source: For $\tau >1/2$ the periodic
orbits must traverse some bonds more than twice. The likelihood of periodic
orbits which traverse the same bonds the same number of times but with
different back-scatter indices $\mu _p$ is increasing, and the interferences
which build $K_{ND}$ are due to the sign correlations among orbits of
exactly the same lengths (when $A\neq 0$ one has to restrict the discussion
to PO's with the same {\it directed} length). This result demonstrates one
important feature of the periodic orbits correlations, namely, that periodic
orbits carry not only metric information (lengths of trajectories) but also
{\it topological information} (Maslov indices and degeneracies). The
distribution of back scatter indices of periodic orbits is a problem that
was not yet addressed by probabilistic graph theory. Our numerical results
together with the general experience from quantum chaos allows us to
conjecture that the spectral form factor connects RMT with the distribution
of back scatter indices on PO's.

Finally, the structure observed in the function $K(\tau)$, decorating the
rather smooth background can be attributed at low $\tau$ to the short and
rather scarce PO. The arrows in Figs.~6a,b indicate their
location. The structures near $\tau =1$ reproduce a trend which was
predicted on different grounds in \cite{keatbog}, namely, the spikes appear
at lengths ${\cal L}- l_p$ (see arrows in Figs.~6a,b). We can explain this
phenomenon with the help of Newton's identities which relate ${\rm tr}
(S(k))^n $ to the coefficients of the characteristic polynomial, and the
inversive symmetry of the latter (see  ~(\ref{ze2})-(\ref{ze5})). Simple
algebra gives
\begin{equation}
\sum_{n=1}^B {\frac{{\rm tr}S^{2B-n}}{2B-n}} = {\rm e}^{i {\cal L} k +
\phi_0} \sum_{n=1}^B {\frac{\left ({\rm tr}S^{n} \right )^* }{n}}\ \ + \cdots
\end{equation}
where the phase $\phi_0$ is independent of $k$ and $\ \cdots \ $ stands for
terms which involve amplitudes and phases of composite orbits. Substituting (%
\ref{posum}) and taking the Fourier transform, we find that the
contributions of the terms ${\rm tr}S^{2B-n}$ to the length spectrum appear
at lengths ${\cal L}- l_p$ where $l_p$ are lengths associated with the
shorter periodic orbits with periods $n$.

When graphs with mixed boundary conditions are investigated, ($\lambda \ne 0$),
the parameter $\Lambda$ which controls the spectral properties depends on the
mean $k_0$ parameter, and a transition from Poisson to RMT statistics is
expected as $k_0$ increases ($\Lambda$ decreases). This transition is
illustrated in Figs.~7a,b, where the dependence of $\tilde{K}(\tau; k_0)$
on $\Lambda$ is displayed.

\subsubsection{\bf Parametric statistics}

So far, we have shown that quantized graphs display most of the generic
statistical properties encountered in the study of ``quantum chaos". We shall
discuss now yet another statistics - the parametric
statistics - and show that the analogy carries over also for this case.
Parametric statistics are defined for systems which depend on an external
parameter (to be denoted by $\chi $), and they give a
quantitative measure for the fluctuations due to level dynamics
\cite{GSBSWZ91}.
Among the first parametric properties studied, were the velocity distribution
${\cal P}(v)$ \cite{veldis} (distribution of the first derivative of the
 levels),
and the curvature distribution ${\cal P}(c)$ (distribution of the second
derivative of the levels)\cite{curvdis}. It has been shown that parametric
statistics are universal for disordered or strongly chaotic
systems, provided the change of $\chi $  does not modify global symmetries.
As is usually the case, non generic classical features may introduce
deviations from the universal parametric statistics.

The parameter which was used to study level dynamics on the graphs
was the bond length of an arbitrarily chosen pair of bonds:
 $L_{i,j} (\chi)= L_{i,j}(0) -\chi$ and $L_{i',j'} (\chi)= L_{i',j'}(0)+\chi$,
so that the total length ${\cal L}$ is  kept constant.  In this way, the
mean density $\bar{d}$ is independent of $\chi$. Moreover, contrary to
the usual studies of parametric statistics, the underlying classical
dynamics of the graph are unaffected by the change of $\chi$ (see also
\cite{KZZ97}).  Modulating the two other parameters of our graphs i.e.
the ``magnetic potential" $A$ and the scattering potential $\lambda$
at each vertex, we are able to study the parametric statistics in the
transition regions where either time reversal symmetry or integrability
are broken.

 To reveal the universality in  ${\cal P}(v)$ and
${\cal P}(c),$ one uses the variance $\sigma _v=\left\langle \left(
\frac{\partial
 k_n}{\partial \chi }\right)^2\right\rangle_k $  to rescale
the velocities $v$ and the curvatures $c$
\begin{equation}
v_n=\frac{\frac{\partial k_n}{\partial \chi }}{\sqrt{\sigma _v}},\medskip\
\,\,\,\,\,\,\,\,\,\,\,\,\,\,\,
c_n=\frac 1{\beta \pi }\frac{\frac{\partial ^2k_n}{\partial \chi ^2}} {
\sigma _v}
 \label{defvel}
\end{equation}
where  $\beta =1,2$ correspond to graphs with or without time reversal
symmetry,
respectively.

The numerical analysis reported below was conducted on a fully connected
hexagon. For intermediate boundary conditions the wavenumber range, over
which we performed our statistical analysis was kept small in order to keep
the control parameter $\Lambda$ to be essentially constant. Then the statistics
was generated over realizations of the lengths of the bonds. By employing a
finite-difference method, we were able to compute the level velocities and the
curvatures for many different values of the parameter $\chi $. The total number
of eigenvalues used to construct the histograms exceeded $186000$ in each
statistics.

We first analyze the velocity distributions for graphs. For level dynamics
within the GOE or the GUE the distribution of level velocities ${\cal P}(v)$
is proved to be Gaussian \cite{veldis}. Some of our results are shown in Fig.
8a,b for $A=0$ and $A\ne 0$, respectively (in both cases, Neumann
boundary conditions where imposed). The calculated velocity distributions
are well approximated by a Gaussian of the same mean value and standard
 deviation.

Fyodorov \cite{F94} derived an analytical formula for ${\cal P}(v)$ which
applies for disordered systems in the strongly localized limit and for which
time reversal symmetry is violated
\begin{equation}
{\cal P}(v)=\frac \pi 6\frac{\pi v\coth \left( \pi v/\sqrt{6}\right) -\sqrt{6%
}}{\sinh ^2(\pi v/\sqrt{6})}.  \label{veldistr}
\end{equation}
 Numerical simulations have shown that this formula is also applicable
in the domain where the Poisson-GOE transition takes place \cite{KZZ97}.
Our numerical results, presented in Fig.~9 show  that  ~(\ref{veldistr})
reproduce the data in the range of large $\Lambda$ values, thus
confirming the suggestion first made in \cite{KZZ97}.

RMT predicts explicit expressions for the curvature distributions
${\cal P}(c)$  \cite{curvdis},\cite{F94}
\begin{equation}
{\cal P}(c)=N_\beta {\frac 1{(1+c^2)^{(\beta +2)/2}}}  \label{curvdistr}
\end{equation}
where $N_\beta $  equals to
$0.5$ and $2/\pi $ for GOE ($\beta=1$) and GUE ($\beta=2$)
respectively. Our numerical calculations for the cases with Neumann
boundary conditions, and $A=0$, $A\ne 0$ are shown in Figs. 10a,b respectively.
The agreement with the theoretical expectation (\ref{curvdistr}) is excellent.

In Ref. \cite{KZZ97}, it was suggested that  ~(\ref{curvdistr})
can be generalized   for the intermediate statistics interpolating between
Poisson and  GOE or GUE. The normalization constant $%
N_\beta $ has to be defined as
\begin{equation}
N_\beta ={\frac 1{\sqrt{\pi }}}{\frac{\Gamma (\frac{\beta +2}2)}{\Gamma (%
\frac{\beta +1}2)}}.  \label{normdis}
\end{equation}
and  $\beta $ takes real values within the interval $
(0,2]$ (see \cite{I90}). We checked this conjecture for our system and we
found that it describes in a satisfactory way the intermediate statistics. For
this, we fitted the tails of the distribution of the second derivative
of the levels (unscaled curvatures) to an algebraic
decay ${\cal P}(C)\sim C^{-(\beta +2)}$. Then the value of $\beta $ found
from the fit was used in  (\ref{defvel}) to rescale the curvatures. Our
results for the case of partially broken time reversal  symmetry
with $A=4$ are shown in Fig.~11a. Similarly, in Fig.~11b we report our
findings for the transition between Poisson and GOE.


\section{\bf Graphs with non uniform connectivity}

So far, we have studied properties of well connected graphs, and
have shown that when the appropriate limit is taken, many statistical
properties of the spectrum reproduce the expectations of RMT. We shall
dedicate the next section to demonstrate cases for which the
connectivity of the graph induces non uniform dynamics  which has
substantial effects on the corresponding quantum spectra and their
statistics.

\subsection {\bf The Hydra }
 As a first example of a family of graphs which is not uniformly connected,
 we studied the ``Hydras" or ``star" graphs. They are graphs which consist of
 $v_0$ bonds, all of which emanate from a single  common vertex labeled
with the
index $i=0$.  The vertex at $i=0$, will be referred to in the sequel as the
Hydra's {\it head}.  The total number of vertices for
such a graph is $V=v_0+1$, and the vertices at the
end of the bonds will be labeled by $i=1 \cdots V$. We shall assume Neumann
 boundary conditions on these vertices. The Hydra
is a bipartite graph, a property which implies e.g., that
there exist no periodic orbits of odd period! This is responsible for
most of the non generic properties of the classical and the quantum properties
of Hydras.

We start with the $S_B-$matrix statistics which will allow us a better
physical understanding of our system. To this end we first calculate the
matrix $T$ defined in   (\ref{scaco}). One can easily show that
\begin{equation}
 T =\left(
\begin{array}{ll}
0 &  I \\
\sigma^{(0)}& 0
\end{array}
\right) \  .\label{starcon}
\end{equation}
The  matrix $\sigma^{(0)}$ is the $v_0\times v_0$ scattering matrix
 at the Hydra's head  as defined in  (\ref{smatrix}).
 $I$ denotes the $v_0\times v_0$ identity matrix. It represents the trivial
back
scattering at the vertices $i=1\cdots V$. The $S_B(k;A)$ matrix  and its square
$S_B^2(k;A) $can be written as
\begin{equation}
 S_B(k;A) =\left( \begin{array}{ll} 0 &  d^{(+)} \\ d^{(-)}\sigma^{(0)}& 0
 \end{array}  \right)
 \  \ \ \ ; \ \ \
S_B^2(k;A) =\left( \begin{array}{ll} d^{(+)}d^{(-)}\sigma^{(0)}
 &  0 \\ 0 &d^{(-)}\sigma^{(0)} d^{(+)}\end{array}  \right)
\label{shydra}
\end{equation}
Where  $d^{(\pm)}_{i,j}= \delta_{i,j} {\rm e}^{ i(k \pm A)L_j } $  are the
diagonal matrices which carry the metric information. It follows from (\ref
{shydra})
that ${\rm tr} S_B^{2n+1} =0$,  and the even traces satisfy ${\rm tr}
S_B^{2n} =
2 {\rm tr} S_H^n $, where  $ S_H \equiv  d^{(-)}\sigma^{(0)} d^{(+)}$.
$S_H$ is a
$v_0\times v_0$ scattering matrix in the space of the Hydra bonds.
It incorporates the reflections from the vertices $i= 1\cdots V$.
$S_H$ is independent of $A$ since all the PO's
on the Hydra are self tracing, and from now on we shall study its properties,
since it is free from the trivial effects which originate from the bipartite
nature of the Hydra.

If we consider the limit $v_0 \gg 1$, and for $n<v_0$, we can use the diagonal
approximation to calculate the form factor of the $S_H$ matrix eigenphase
spectrum. In the present context, the classical evolution operator which
corresponds to the quantum $S_H$ matrix is
\begin{equation}
\left (U_H\right )_{i,j } =   {2\over v_0^2} (1+\cos\omega_0 )  +
 \left ( 1-{  2\over v_0}(1+\cos \omega_0)  \right )
\delta_{i,j} \ .
\end{equation}
The spectrum of $U_H$ consists of the values $1$ and
$1-2{ 1+\cos\omega_0 \over v_0}$ which is $v_0-1$ times degenerate. Therefore,
\begin{equation}
\label{startrU}
u_n \equiv {\rm tr} U_H^n = 1+ (v_0-1)\left (1-{2\over v_0}( 1+\cos\omega_0)
\right)^n \ .
\end{equation}
For large valencies $v_0$ the ${\rm tr}S_H^n$ are dominated by the $n$
 repetitions
of fixed points. Thus, we can write for the trace of $S_H^n$
\begin{equation}
{\rm tr}S_H^n=\sum_{j=1}^{v_0}e^{2ikL_jn}\left( \frac{1+e^{-i\omega _0}}{v_0
}-1\right) ^n+\sum_pn_pg_p{\cal A}_p^re^{il_pk}  \label{starpff1}
\end{equation}
where the second sum contains contributions from other periodic orbits. By
performing furthermore an average over realizations of the lengths of the
bonds of Hydras we get
\begin{equation}
\left\langle \left| {\rm tr}S_H^n\right| ^2\right\rangle =v_0\left| \left(
\frac{1+e^{-i\omega _0}}{v_0}-1\right) ^n\right| ^2+n\left[ u^n-v_0
\left(\frac 2{v^2}(1+cos\omega_0) - \frac 2v_0(1+cos\omega_0)+1\right)^n
\right].
\label{starpff2}
\end{equation}
In the second term of the r.h.s of   (\ref{starpff2}) we have subtracted
from the classical return probabilities $u_n$ (see   (\ref{startrU}),
the contributions from the fixed points which already had been taken into
 account.
  (\ref{starpff2}) can be written in a better form i.e.
\begin{eqnarray}
\frac 1{v_0}\left\langle \left| {\rm tr}S_H^n\right| ^2\right\rangle
&=&\left[ \frac{2\left(1-v_0\right) \left( 1+\cos \omega _0\right) +v_0^2}{
v_0^2}\right] ^n+\frac n{v_0}\left[ \left( v_0-1\right) \left( 1-2\frac{
1+\cos \omega _0}{v_0}\right) ^n+1\right] -  \label{starfinal} \\
&&n\left[ \frac 2{v_0^2}\left( 1+\cos \omega _0\right) -\frac 2{v_0}\left(
1+\cos \omega _0\right) +1\right] ^n.  \nonumber
\end{eqnarray}
 This  is different from the generic expression in two important ways: The
linear term which dominates the larger $n$ domain is proportional to $n$ and
not to $2n$ because in the Hydra all the periodic orbits are self  tracing.
The other two terms which dominates the domain of smaller $n$, are due to the
$n$ repetitions of fixed points and to small (degenerate) eigenvalues of $U_H$.
Their $n$ dependence cannot be scaled with $v_0$, and they represents a typical
transient effect. The spectrum of $S_H$ is not degenerate, and therefore for
$n>v_0$ the form factor reaches its asymptotic value $1$. This transition is
due to the interference of contributions of periodic orbits with the same
lengths, but even for the simple  Hydra graphs we do not have a theory which
explains this phenomenon.

The above approximation applies quite well also for the Dirichlet boundary
conditions. Indeed, for $\omega_0=\pi$ we have from (\ref {starfinal}) that
$\frac 1{v_0}\left\langle \left| {\rm tr}S_H^n\right| ^2\right\rangle = 1$.
Moreover, the transition to the Poisson limit is again described quite well by
(\ref {starfinal} ), for $\tau \equiv {n\over v_0} <1$. This is shown in
Fig. 12
for a Hydra with $v_0=50$ and various values of the parameter $\Lambda _0.$
The non generic features expected from the periodic orbit theory are well
reproduced by the numerical data.

The wavenumber spectrum follows the same trends as the eigenphase spectrum
discussed above.  The secular equation  (\ref{meq}) for  Hydras takes a
particularly simple form,
\begin{equation}
\sum_{j=1}^{v_0} \tan kL_j = \frac{\lambda _0}k.   \label{starsecu}
\end{equation}
Here, we investigate for simplicity the case of Hydras with a zero scattering
potential $ \lambda _0=0$. In Fig.~13 we present the level spacing distribution
$P(s)$ for various valencies. In all cases, the spectrum shows level repulsion,
which is described rather well by the Wigner surmise for large valencies. A
closer look shows deviations which do not decrease when the valency increases.
In particular, in the limit $s\rightarrow 0$, the $I(s)$ does not approach the
expected power law, and this can be seen in the inset of Fig.~13 where the
 results for
Hydras with $v_0=7-18$ are compared with the GOE expression. In light of the
discussion of the $S_H$ eigenvalues statistics, the deviations observed in the
nearest neighbor spacings distribution are not surprising.

In a previous publication \cite{KS97} we investigated the two point form factor
$K(\tau )$ for the Hydras and we had reported (for the case $v_0=5$) quite good
agreement with RMT predictions. However, upon increasing $v_0$, $K(\tau )$
shows
the same behavior as its counterpart for the eigenphase spectrum of $S_H$. In
Fig.~14 we present the two-point form factor for Hydras with $v_0=5,15$. The
deviations from RMT predictions, as we are increasing the valency, follow the
pattern observed for the $S_H$ spectral statistics (see   (\ref{starfinal})).

As expected, the Hydra spectrum displays non generic spectral statistics,
which reflect the special connectivity and hence the classical dynamics of this
graph.  Periodic orbits theory (in particular, its diagonal approximation)
 reproduces
the statistics which relate to the limit of short evolution times.


\subsection{\bf Rings With Variable Connectivity}

In this section we shall study a family of regular graphs which are
defined by the connectivity matrix
\begin{equation}
C_{i,j} =\left \{
 \begin{array} {ll} 1   & {\rm for}\  i\ne j\  {\rm and}\  |i-j| \le b  \\
0 & {\rm for}\  i= j\  {\rm or}\ \ \  |i-j| >b
 \end {array} \right.
\end{equation}
so that $2b$ is the valency. When the graph is drawn in the plane, with the
$V$ vertices placed  on a circle with $\theta_i >\theta_j\  {\rm for}\  i>j$,
and $V\gg b$, then the resulting shape is a ring, and hence the name.

 We shall study the  spectral statistics as a function of
the valency, and will attempt to answer the question, at what value of
$b$ can we consider the graph as sufficiently well connected so that
its spectral properties can be reproduced by RMT. We shall assume
Neumann boundary conditions throughout this section.

  As long as  $1 <b \ll V$, the classical evolution operator
describes  a random walker on a ring, where the hopping step size can
take any value between $1$ and $b$. This results in a diffusive evolution
on the ring, with $D \approx {2\over 3} \pi^2 b^2$. The case  $b=1$ is
trivial because the Neumann boundary conditions in this case does not
allow reflections at the vertices, and hence, the particle goes around
the ring ``ballistically". To eliminate this effect we added loops
at each vertex. This does not alter the diffusive dynamics, but allows
us to include rings with nearest neighbor hopping.

The quantum dynamics is strongly affected by the fact that the lengths of the
bonds are rationally independent. In the limit $b \ll V$ this causes the
eigenstates to be localized, and the spectral correlations will bear the marks
of the degree of localization. Our purpose is to study the transition in
the spectral statistics of rings, as $b$ is increased. The addition of loops
to the vertices, introduces also a subspace of eigenfunctions for which the
vertex values $\phi_i=0$, and the wavefunction itself vanishes on all the bonds
and all the loops but one. There it takes the form $\sin {\pi n \over
L^{(l)}_i} x$,
where  $L^{(l)}_i$ is the length of the loop attached at the $i$ vertex. This
subspace of eigenstates and the corresponding spectrum will be excluded
from the
following statistical study.

Starting with rings with $b=1$, we can calculate their properties by
introducing
the quantum transfer operator
\begin{equation}
\label{transf1}
T_N = \prod_{i=1}^{V} T_i\ \ \ \ \ ;\,\,\,\,\,\
T_i =
\left( \begin{array}{ll}
 1-\rho(k,A^{(l)}_i)  &
 -\rho(k,A^{(l)}_i) \\
\rho(k,A^{(l)}_i)       &
1+\rho(k,A^{(l)}_i)
\end{array}\right)\cdot
\left( \begin{array}{ll}
{\rm e}^{i(k+A_i)L_i}  &
 0 \\
0     & {\rm e}^{i(-k+A_i)L_i}
\end{array}\right)
\end{equation}
where the superscript $(l)$ distinguishes the loops parameters. The bond
$(i-1,i)$ is referred to as the  $i$ bond. The parameter $\rho$ is given
by
\begin {equation}
\rho(k,A^{(l)}_i) = 2i {\sin{A^{(l)}_i+k\over 2}L_i^{(l)}
\sin{A^{(l)}_i-k\over 2}L_i^{(l)}\over \sin kL_i^{(l)}} \ .
\end{equation}

By multiplying the transfer matrices for arbitrarily long segments, we
calculated the Lyapunov exponent which yields the localization length
$l_{\infty}$. Since for all $k$ values we found it to be of order $1$,
we expect that the eigenfunctions for rings with $V>10$ will be
well localized. A direct inspection of the eigenstates confirms this
expectation (see Fig.~15).

The effect of localization on the spectral statistics can be understood
by the following argument. As long as
the total length $\cal{L}$ of the ring is sufficiently larger than the
Anderson localization length $l_{\infty }$, one can approximate the spectrum
qualitatively as a union of ${\cal{L}}/l_{\infty}$ uncorrelated spectra.
One expects that the form factor for two  rings with lengths
${\cal{L}}$ and ${\cal{L}}'$ will be related  by \cite{P81}
\begin{equation}
\label{uncff}
K_{{\cal{L}}}(\tau) \simeq K_{{\cal{L}}'} (\tau {\cal{L}}/{\cal{L}}').
\end{equation}
  In Fig.~16 we plot the integrated form factor $\tilde{K}(\tau )$ (\ref{iff})
for $b=1$ graphs with number of vertices $V=11$
and $V=22$. Scaling the $\tau$ axis by 2, the two form factors coincide
as expected. The relation (\ref{uncff}) provides the correct sense in which
one should interpret the statement that the spectrum of a localized system
tends to the Poissonian limit as the size increases. The correlations in the
spectrum remain, but they are on a scale which is ${\cal{L}}/l_{\infty}$ larger
than the mean spacing.

 As the connectivity is increased, and keeping the length constant, one
expects to see a transition in the spectral statistics. The first statistical
quantity that we investigate is the integrated level spacing distribution
$I(s)$. Our results for the case $V=22$ are shown in Fig.~17. As the
connectivity range increases, $I(s)$ makes a transition from the Poisson
distribution towards the GOE. Recently, it was suggested \cite {bogomo} that
there exists a universal intermediate statistics, which applies for systems
which undergo a transition from Poisson to RMT statistics. At the intermediate
(critical) point,
\begin{equation}
P_{cr}(s)=4s\exp (-2s).  \label{bog1}
\end{equation}
Fig.~17 shows an excellent agreement between our numerical results for the
ring with $b=2$ and  ~(\ref{bog1}). Similar degree of agreement with the
critical statistics appears also for other statistical measures like the
number variance. The number variance $\Sigma ^2(L)$ probes the spectrum
over all
correlation lengths $L$ and describes the fluctuations of the number $n(L)$ of
levels contained in a randomly chosen interval of length $L$. Is defined as
\begin{equation}
\label{nuvar}
\Sigma ^2(L)\equiv \left\langle (n(L)-L)^2\right\rangle
\end{equation}
where the angular brackets $\left\langle {}\right\rangle $ denote a local
averaging over sufficiently many levels. $\Sigma ^2(L)$ is related with $K(\tau
 )$
by the integral transform
\begin{equation}  \label{2sigma}
\Sigma^2 (L)= \frac 2\pi \int_0^{\infty} d\tau \frac {sin^2(\pi L\tau)}{
\tau^2} K(\tau ;\bar{k}).
\end{equation}
and the reason that we choose here to concentrate on $\Sigma ^2(L)$ is that it
 behaves
quite nice, with respect to the highly fluctuative form factor.

The general expectation for generic systems \cite{berry}, is that $\Sigma
^2(L)$
should comply with the predictions of RMT for small values of $L$ (universal
regime) and saturate to a non-universal value for large $L$'s due to the
semiclassical contributions of short periodic orbits. Our results are presented
in Fig.~18 together with the critical number variance which reads \cite{bogomo}
\begin{equation}
\label{bogo2}
\Sigma^2_{cr}(L)=\frac L2 + \frac{1-exp(-4L)}{8}.
\end{equation}
For reference we also draw the number variance for a Poissonian spectrum given
by $\Sigma _2(L)=L$, and the RMT expectation
\begin{eqnarray}
\Sigma _2(L)_{GOE} &=&\text{ }\frac 2{\pi ^2}\{\ln \left( 2\pi L\right)
+\gamma +1+\frac 12 {\rm Si}^2(\pi L)-\frac \pi 2%
\mathop{\rm Si}
(\pi L)-\cos (2\pi L)-  \label{sigmagoe} \\
&&
\mathop{\rm Ci}
(2\pi L)+\pi ^2L\left( 1-\frac 2\pi
\mathop{\rm Si}
(2\pi L)\right) \}  \nonumber
\end{eqnarray}
where $\gamma $ is the Euler constant.

As one increases the connectivity the graph statistics approach the  RMT
expressions. On the basis of the estimate of the diffusion constant on the
ring, we expect that the RMT limit is reached for $b \simeq V$.

\section{\bf Conclusions}

 The graph is a one dimensional system. Yet, it is not simply connected, and
this is why it can display chaotic classical dynamics, in the sense explained
above.  The fact that the graph is one dimensional can be seen in various
classical and quantal attributes. On the quantum level, the smooth spectral
counting function is proportional in general to $E^{d/2}$ and indeed, in
our case
$d=1$. The stability factors in the trace formula correspond to a system with
a single expanding direction, and no contracting direction. In this respect
the graph trace formula is similar to the Riemann-Weyl trace formula. (The
counting function for the Riemann zeros, however, has a logarithmic correction
to the strictly one dimensional Weyl term).

 In spite of the simplicity of the graph Schroedinger operator,
 we have shown that the spectrum displays many features which appear
in the study of the quantum analogues of classically chaotic systems. The
limit $\hbar \rightarrow \infty$ can be replaced by the limit
${\cal L}\rightarrow \infty$, which ensures that all the features
which are due to the short, and non generic  periodic orbits are of
lesser relative importance.

For a well connected graph, the length spectra of periodic orbits of
periods larger
than $V$ show increasing degree of degeneracy as the period increases.
 These  degeneracies give rise to the correct
behavior of the form factor at large values of $\tau$. In other words, the
fact that the quantum spectrum is real and discrete, is expressed in the trace
formula through the degeneracies of periodic orbit lengths, and the
distribution
of the corresponding back scattering indices. This observation points at a
possible
new interpretation of the Wigner Dyson theory in terms of probabilistic graph
theory.

 The results reported here encourage us to believe that quantum graphs may
serve
as a convenient tool for the study of quantum  chaotic or disordered
systems. We do not know yet if it is possible to map any given Hamiltonian
system
onto a corresponding graph. The evidence is mounting, however, in favor of
a very intimate link, and the search for this connection is one of our
immediate goals.

\section{\bf Acknowledgments}

This research was supported by the Minerva Center for Physics of Nonlinear
Systems, and by a grant from the Israel Science Foundation. (T.\thinspace
K.) thanks the Feinberg School at the WIS for a Post Doctoral fellowship.
We thank D. Cohen, D.  Miller, H. Primack, Z. Rudnick and H. Schanz for
valuable suggestions and comments, and Y. Colin De Verdiere for bringing
J. P. Roth's paper \cite{R83} to our attention. In particular we thank
U. Gavish for reading the manuscript critically and making valuable comments.

\newpage

\begin{figure}
\psfig{figure=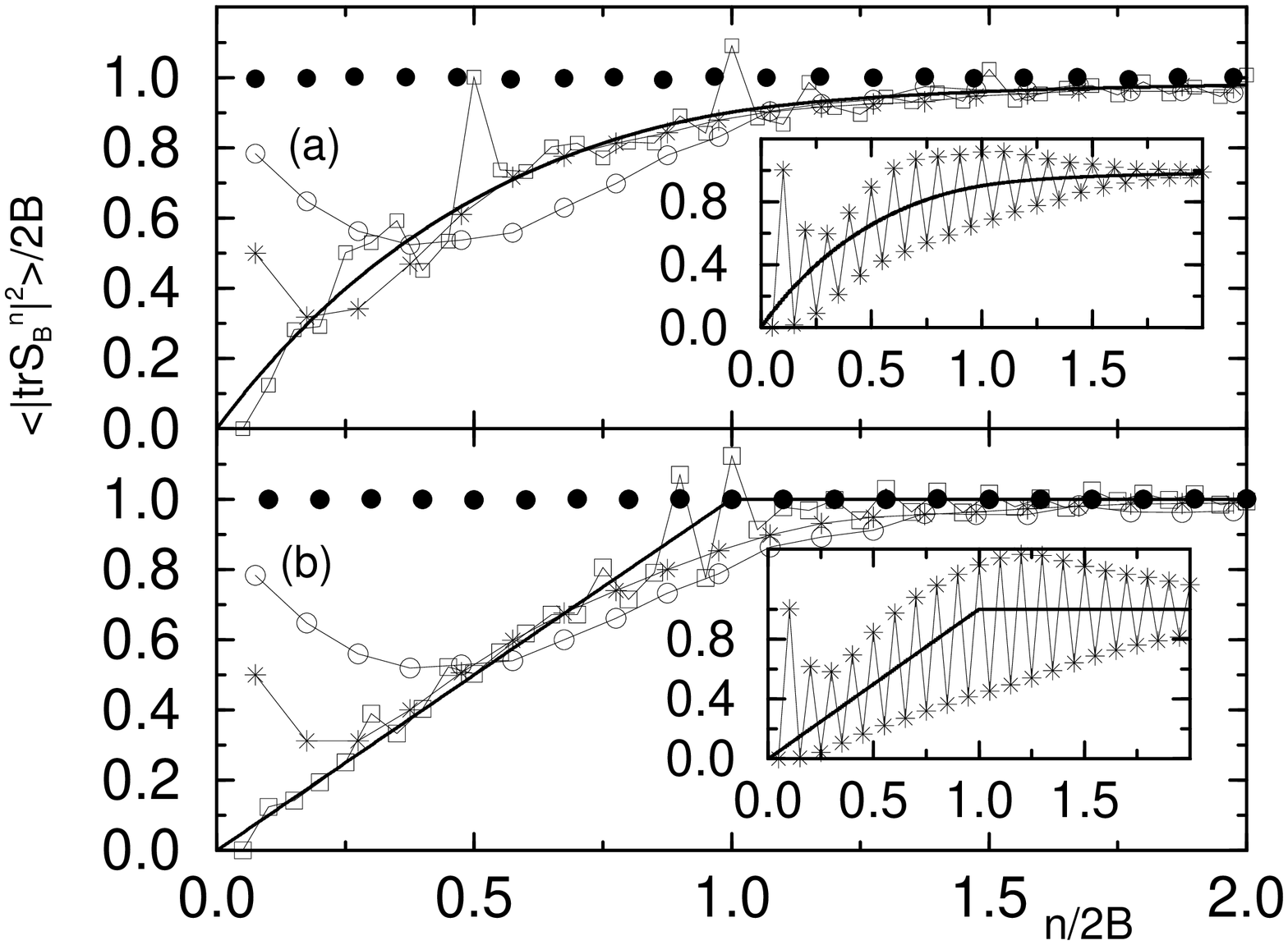,height=8cm,width=16cm,angle=270}
Fig.~1: The form factor for the eigen-phase spectrum of $S_B$
for a pentagon graph. Bold solid lines are the expectations
for the COE and CUE expressions.  The data
are averaged over odd-even powers of the form factor as explained
in the text. In the insets we present the form factor for the
case $\Lambda=0.833$ ($\circ$) without averaging.
\\
(a) $A=0$, $\Lambda = 0$ ($\Box$), $\Lambda = 1.25$ ($\ast$),
$\Lambda = 2.5$ ($\circ$)and $\Lambda = 3500$ ($\bullet$).\\
(b) $A\neq 0$ with the same boundary conditions as in (a)\\
\end{figure}

\begin{figure}
\psfig{figure=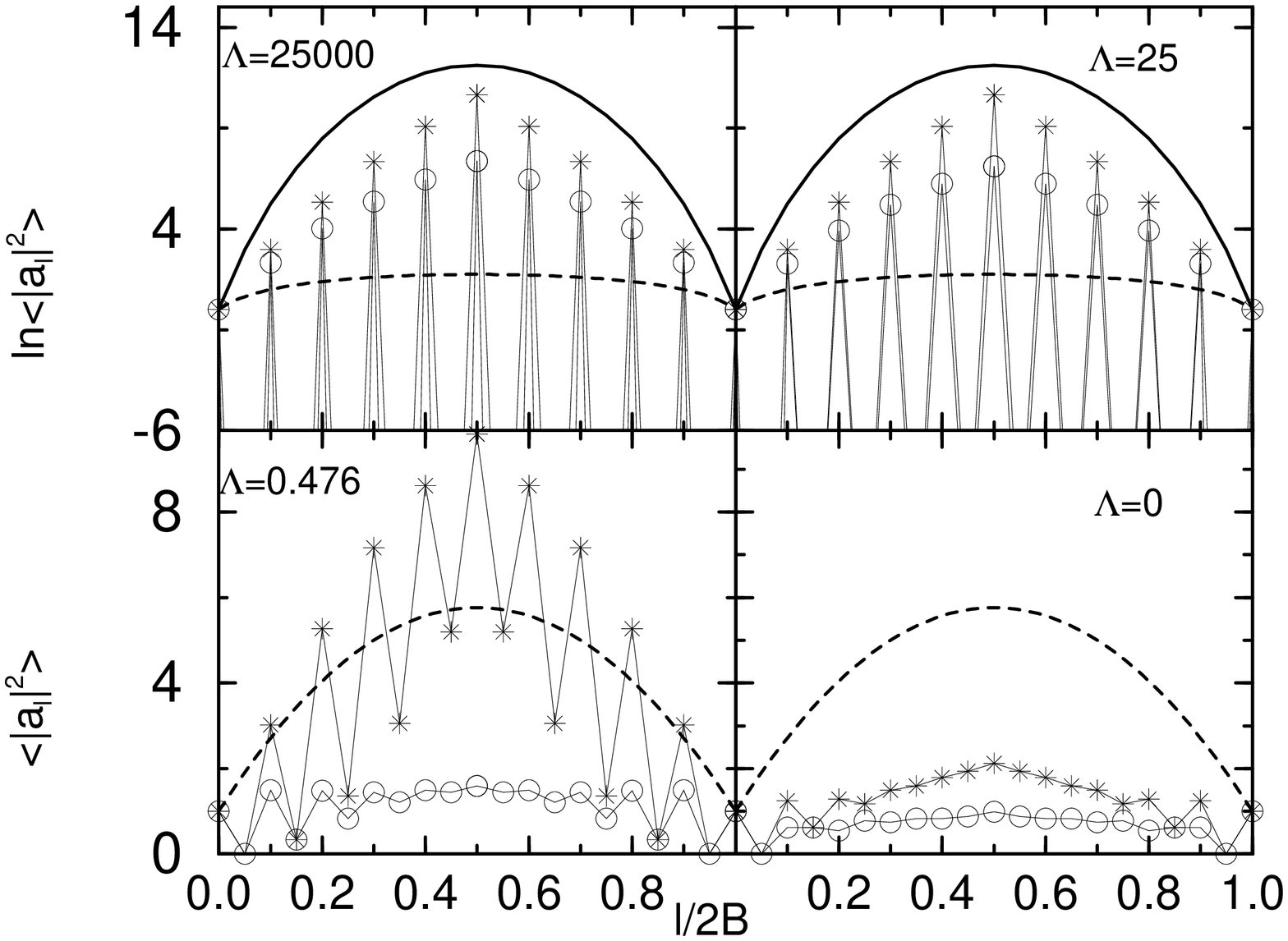,height=8cm,width=16cm,angle=270}
\psfig{figure=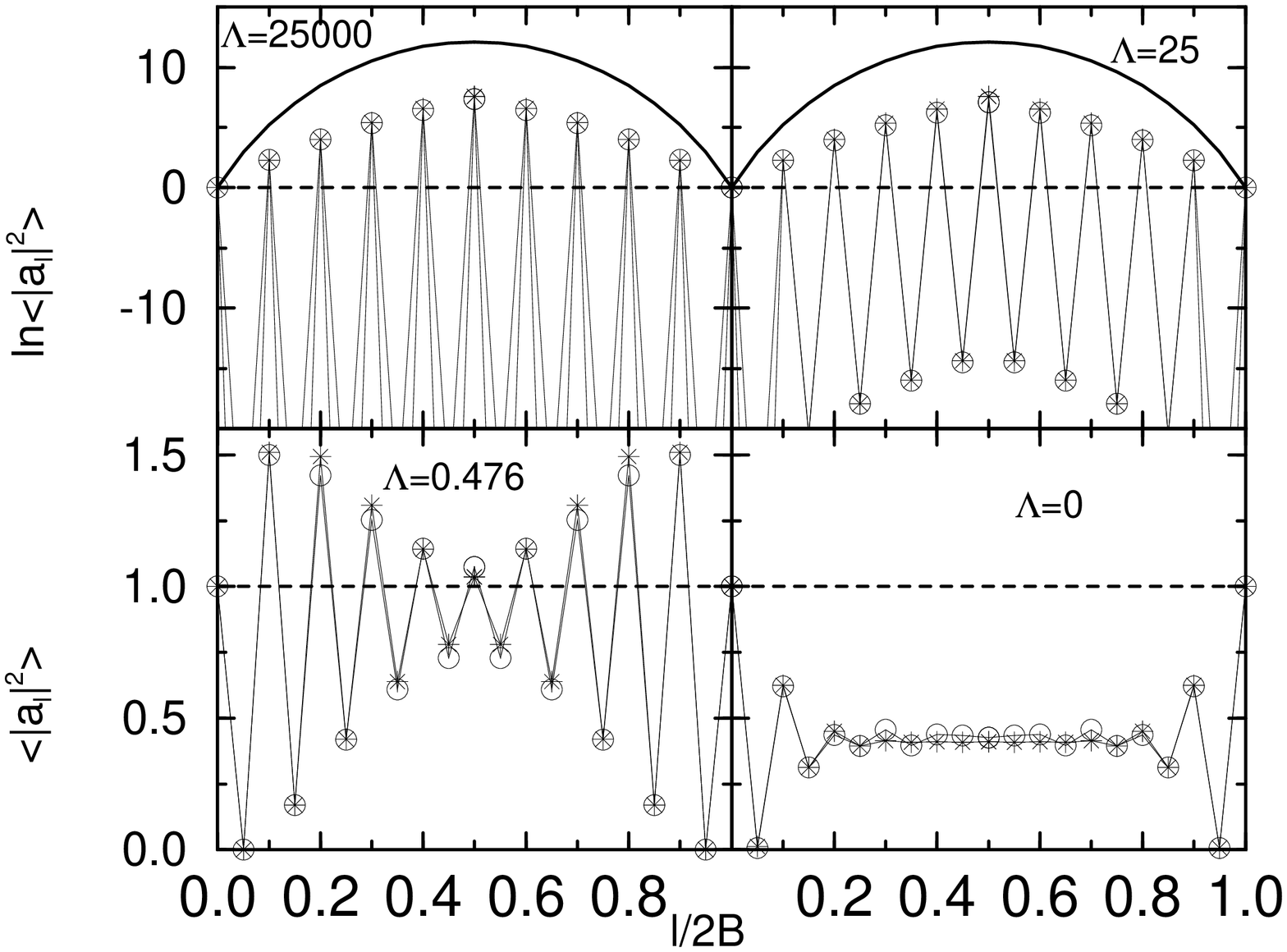,height=8cm,width=16cm,angle=270}
Fig.~2: The Fourier components $\left<|a_l|^2\right>$ of $C_{\zeta}$
for a pentagon graph. ($\ast$) corresponds to semi-classical calculations
while ($\circ$) to the exact quantum mechanical calculations. \\
(a) $A=0$, $\Lambda=25000$, $\Lambda=25$, $\Lambda = 0.476$ and
$\Lambda = 0$;\\
(b) $A\neq 0$, with the same boundary conditions.\\
In both figures the bold solid line correspond to Poisson while the
bold dashed line is the expectation of the corresponding RMT.
\end{figure}

\begin{figure}
\psfig{figure=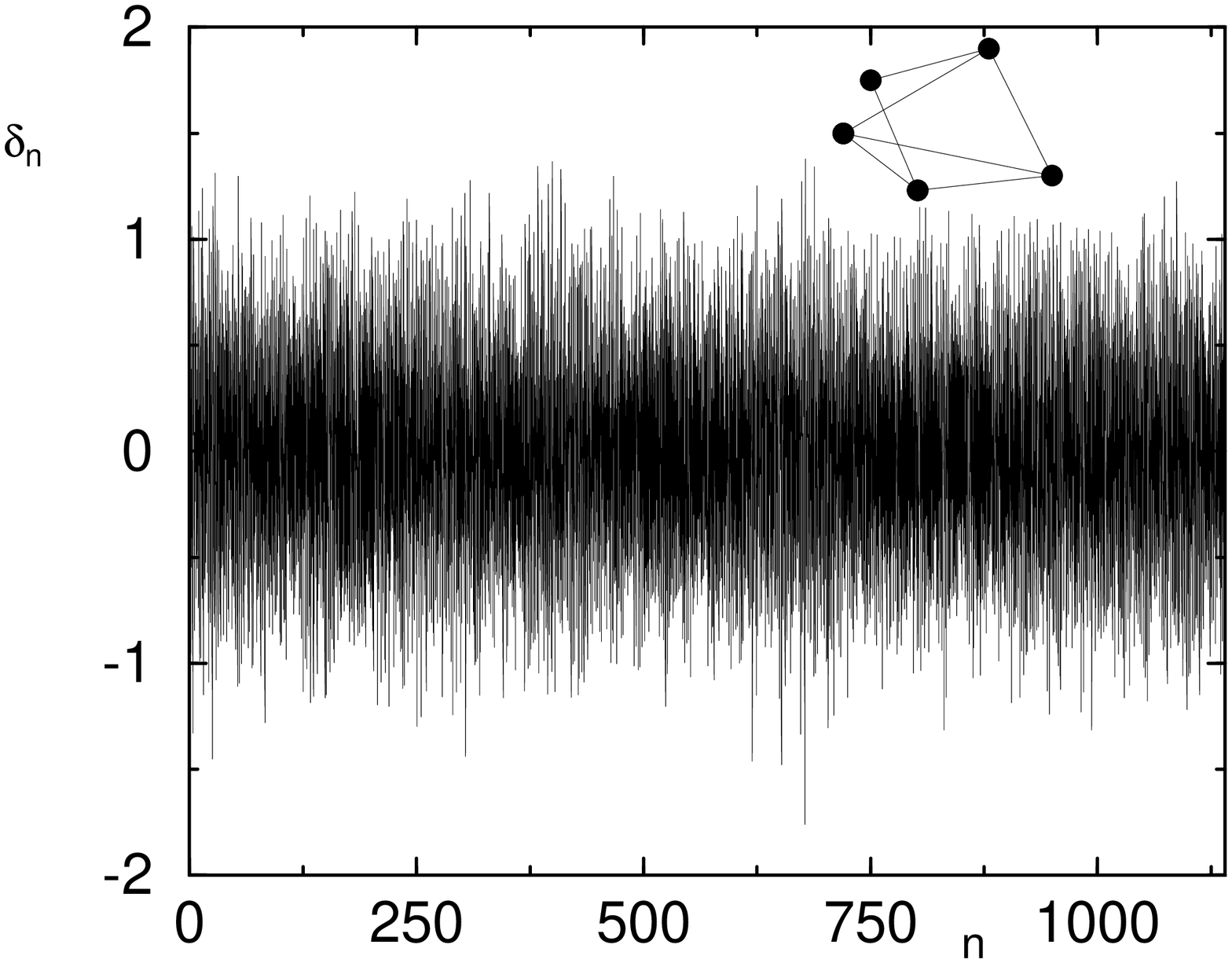,height=8cm,width=16cm,angle=270}
Fig.~3: $\delta_n = N(k) - {\bar N}(k)$ vs. the wavenumber label $n$
for the graph shown in the inset.
\end{figure}

\begin{figure}
\psfig{figure=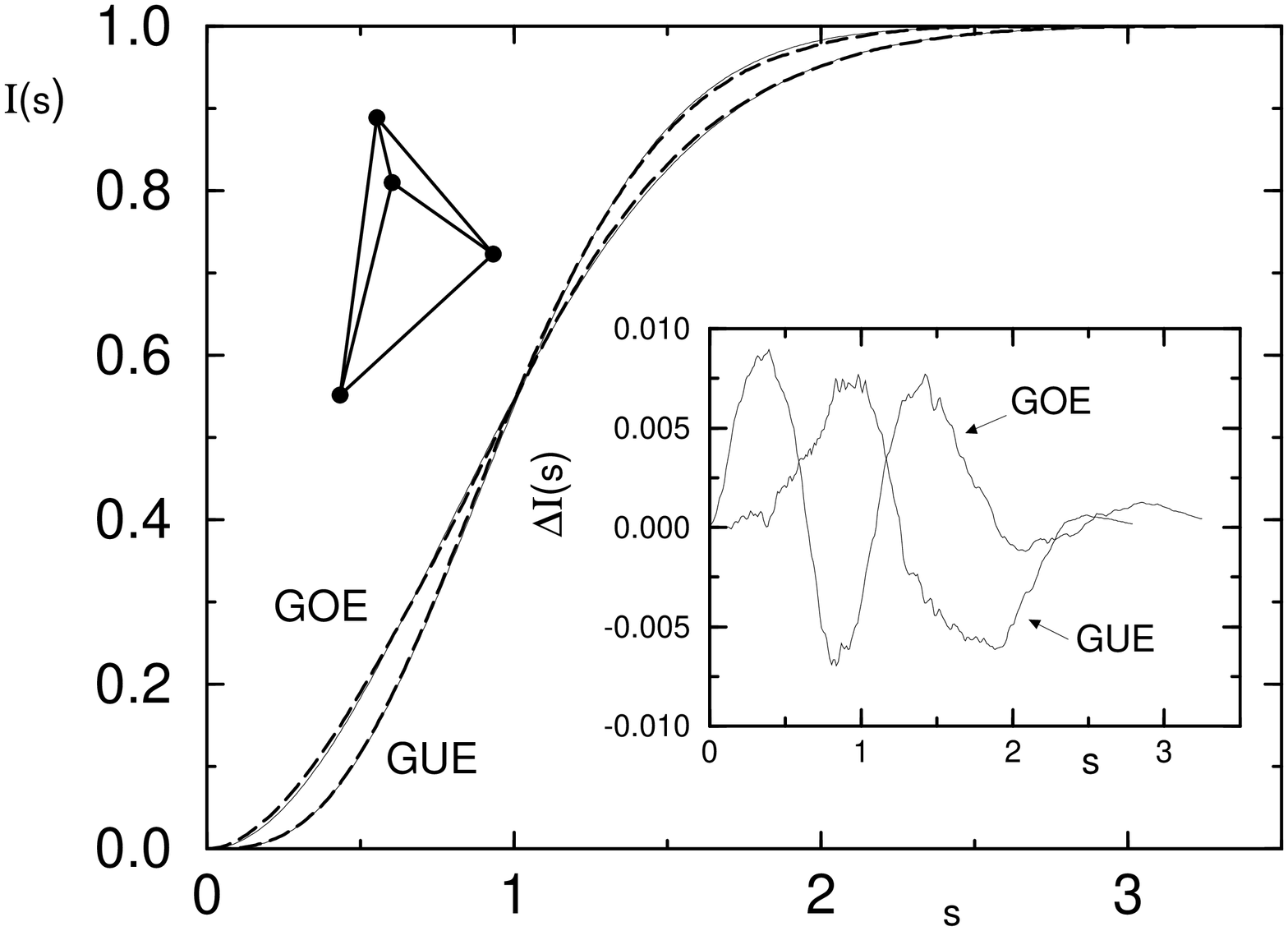,height=8cm,width=16cm,angle=270}
Fig.~4: Integrated nearest neighbor distribution $I(s)$ for a fully
connected quadrangle with $\Lambda= 0$ (Neumann boundary conditions). The
results are based on the lowest $80,000$ levels of a single realization
of the bonds. $\Delta I$ indicates the deviation from the exact GOE/GUE
results;\\
\end{figure}

\begin{figure}
\psfig{figure=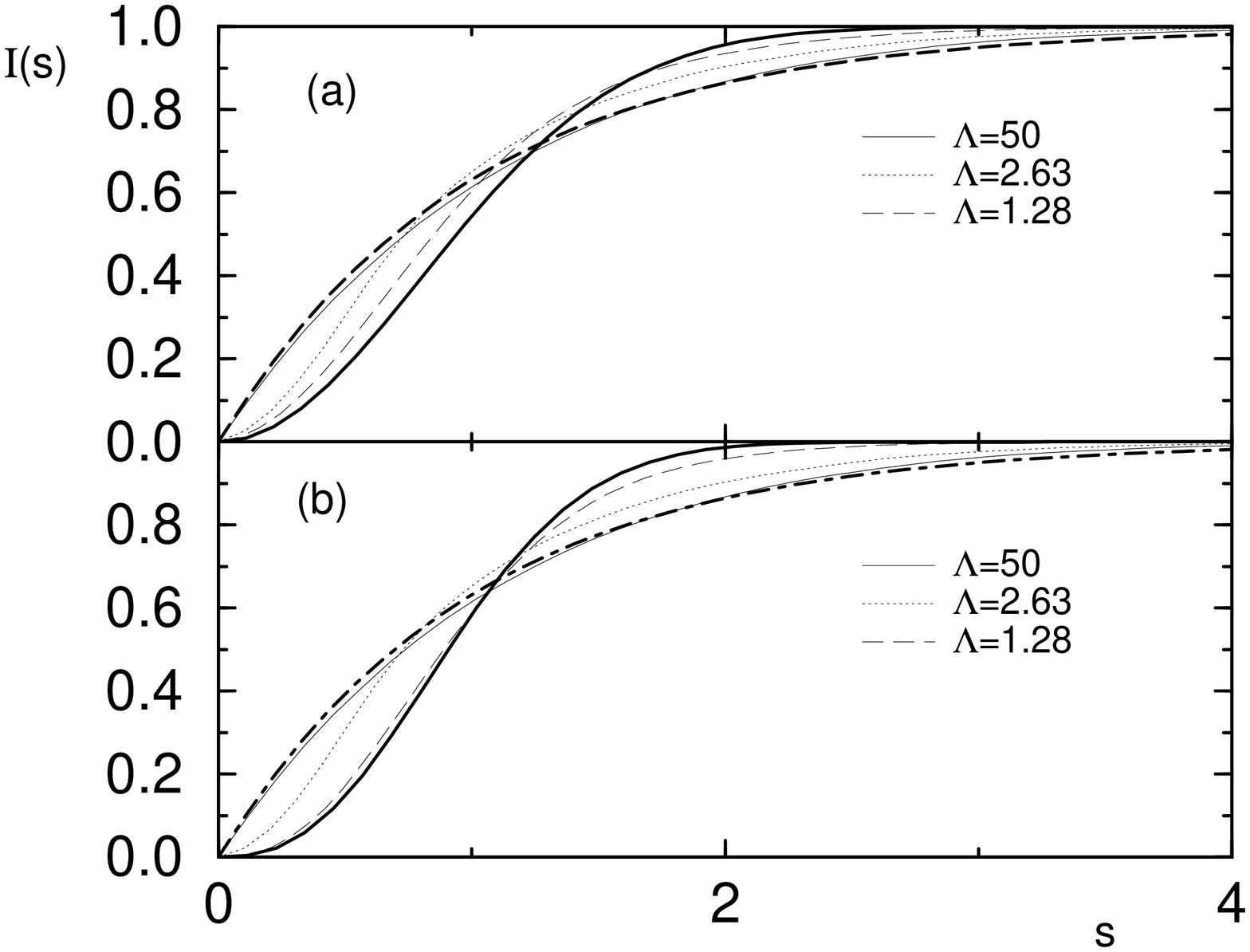,height=8cm,width=16cm,angle=270}
Fig.~5: The integrated nearest neighbor distribution $I(s)$ for a fully
connected pentagon and various values of the parameter $\Lambda$.
The statistics was generated over a large number of realizations
of the bonds of the graph;\\
(a) $A=0$ and various values of the parameter $\Lambda$.
(b) $A\neq 0$ with the same boundary conditions as in (b).\\
In both figures the thick solid line is the expectation of the
corresponding RMT while the thick dashed line correspond to Poisson .
\end{figure}

\begin{figure}
\psfig{figure=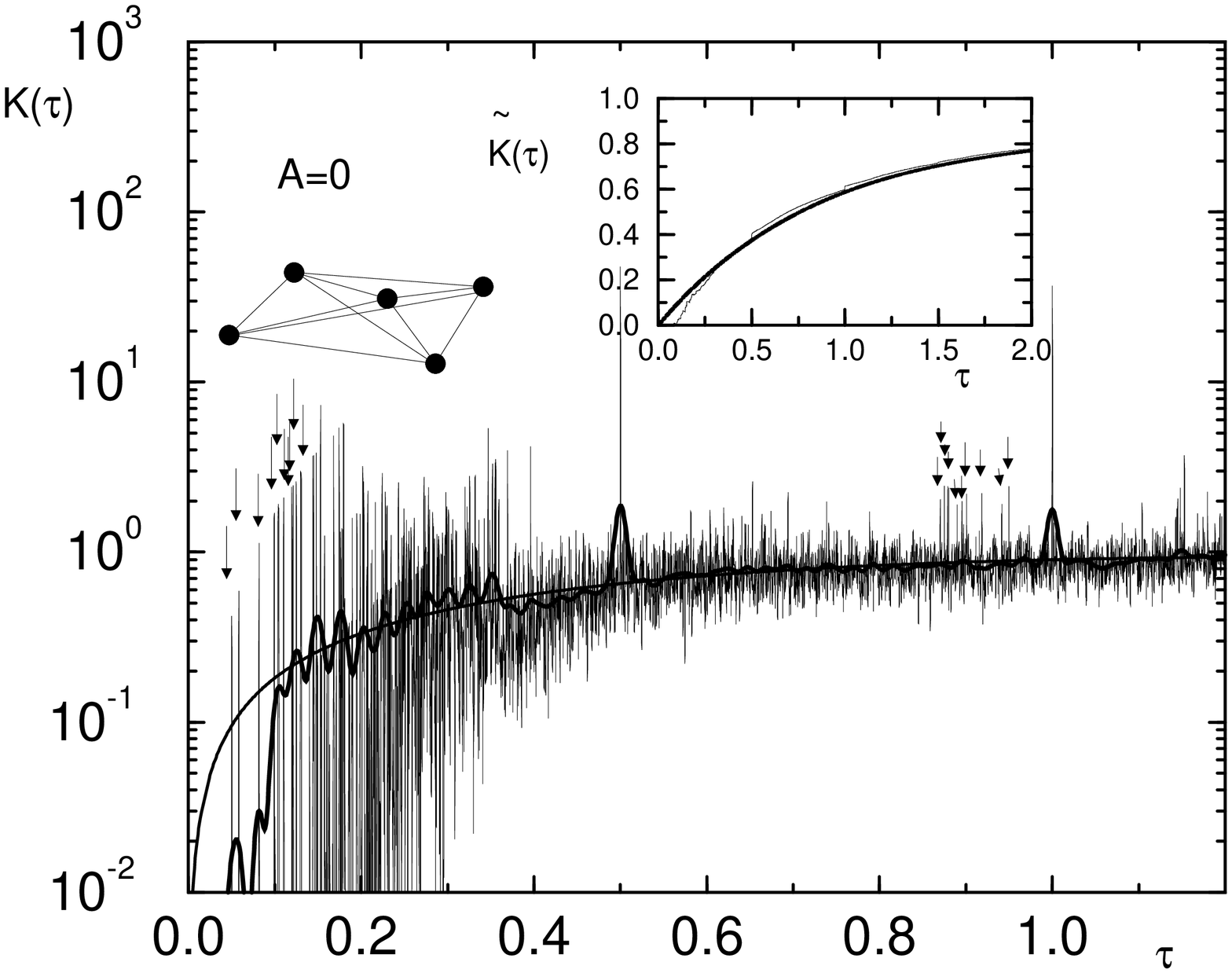,height=8cm,width=16cm,angle=270}
\psfig{figure=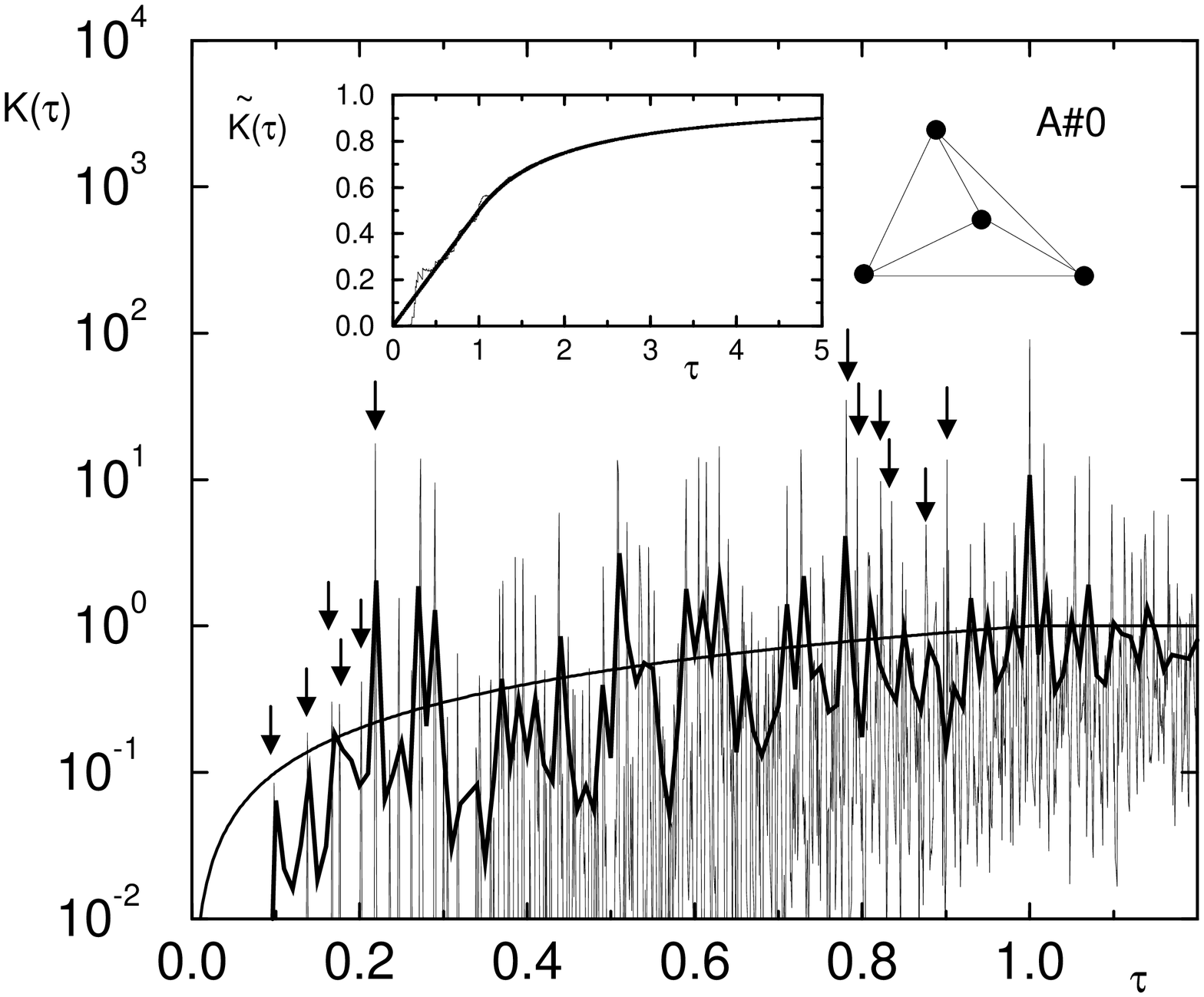,height=8cm,width=16cm,angle=270}
Fig.~6: Two point form factor ($\approx 100,000$ levels). The arrows
indicates the location of the short periodic orbits and their reciprocal
lengths with respect to the Heisenberg length. In the insets we show
the corresponding integrated form factor $\tilde{K} (\tau )$ (thin line).\\
(a) Fully connected pentagon with $A=0$;\\
(b) Fully connected quadrangle and $A\neq 0$.\\
In both figures the bold lines are the expectation of the corresponding RMT.
\end{figure}

\begin{figure}
\psfig{figure=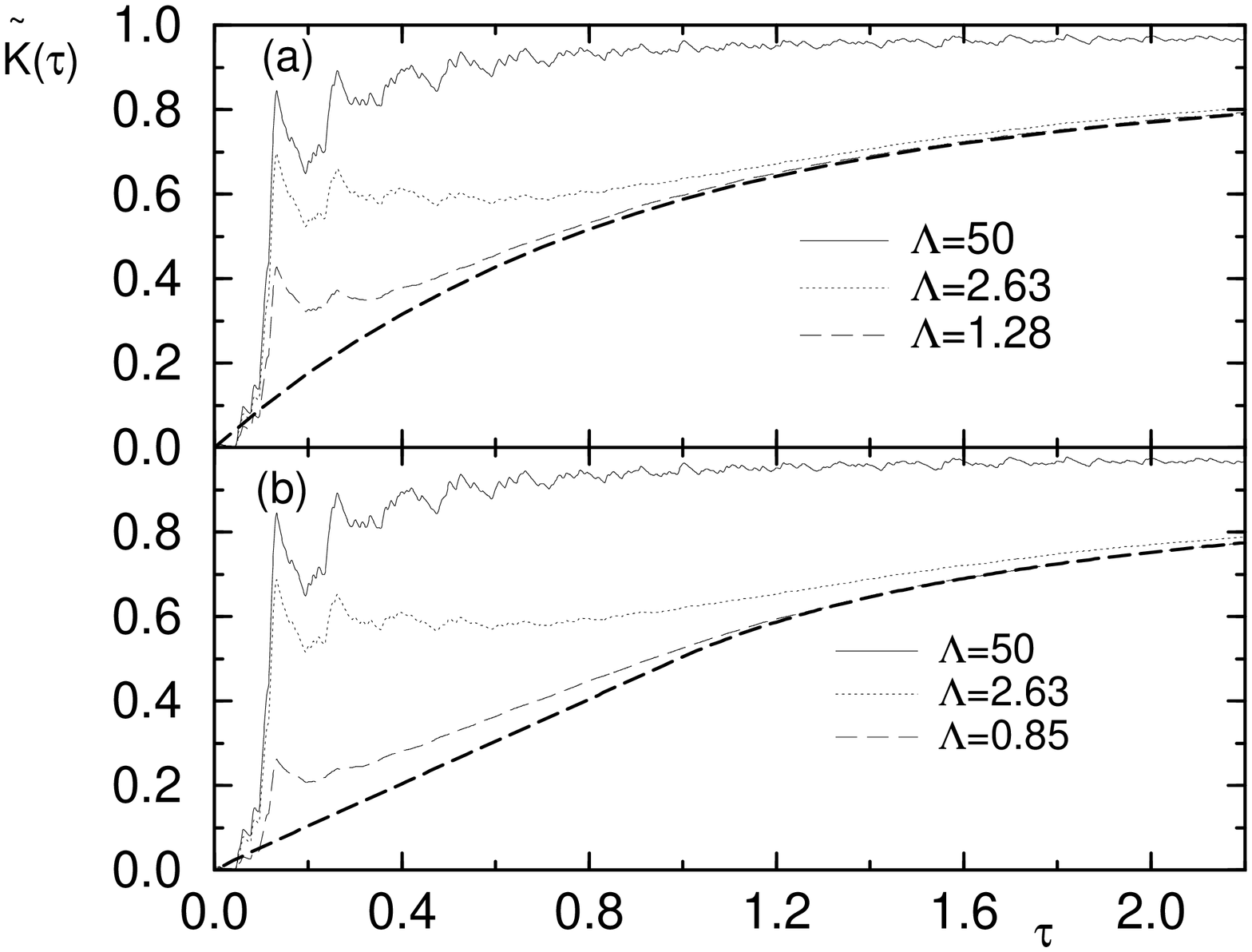,height=8cm,width=16cm,angle=270}
Fig.~7: The integrated two point form factor $\tilde{K} (\tau)$ for a fully
connected pentagon. The thick dashed lines correspond to the RMT
expectations.\\
(a) $A=0$ and various values of $\Lambda$.\\
(b) $A\neq 0$ and various values of $\Lambda$.\\
\end{figure}

\begin{figure}
\psfig{figure=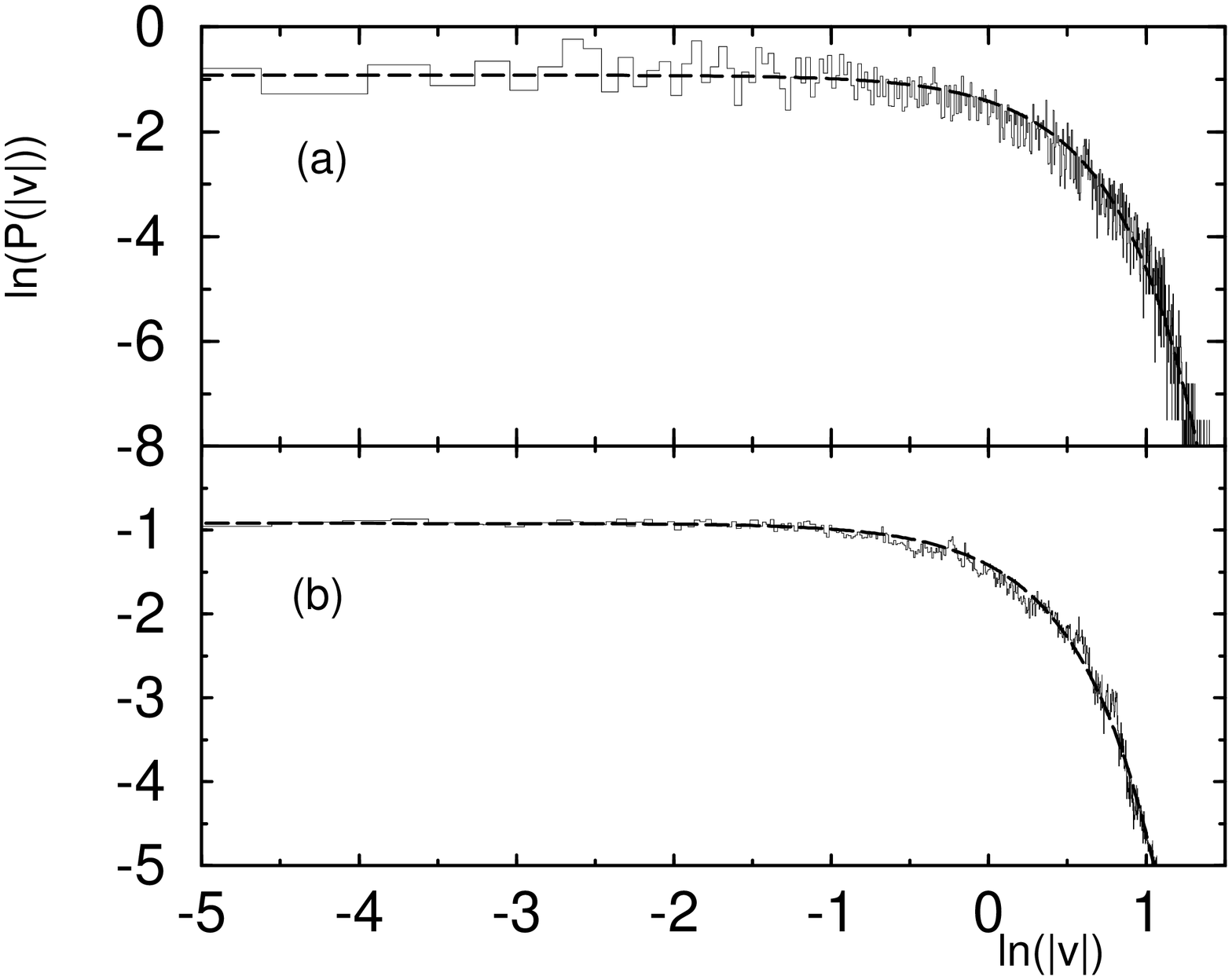,height=8cm,width=16cm,angle=270}
Fig.~8: Velocity distribution $P(v)$ for a fully connected hexagon. The thick
dashed line correspond to a Gaussian with the same mean and standard
deviation, with the numerical data;\\
(a) $A=0$, $\Lambda_i = 0$ (Neumann boundary conditions).\\
(b) $A\neq 0$, $\Lambda_i = 0$.\\
\end{figure}

\begin{figure}
\psfig{figure=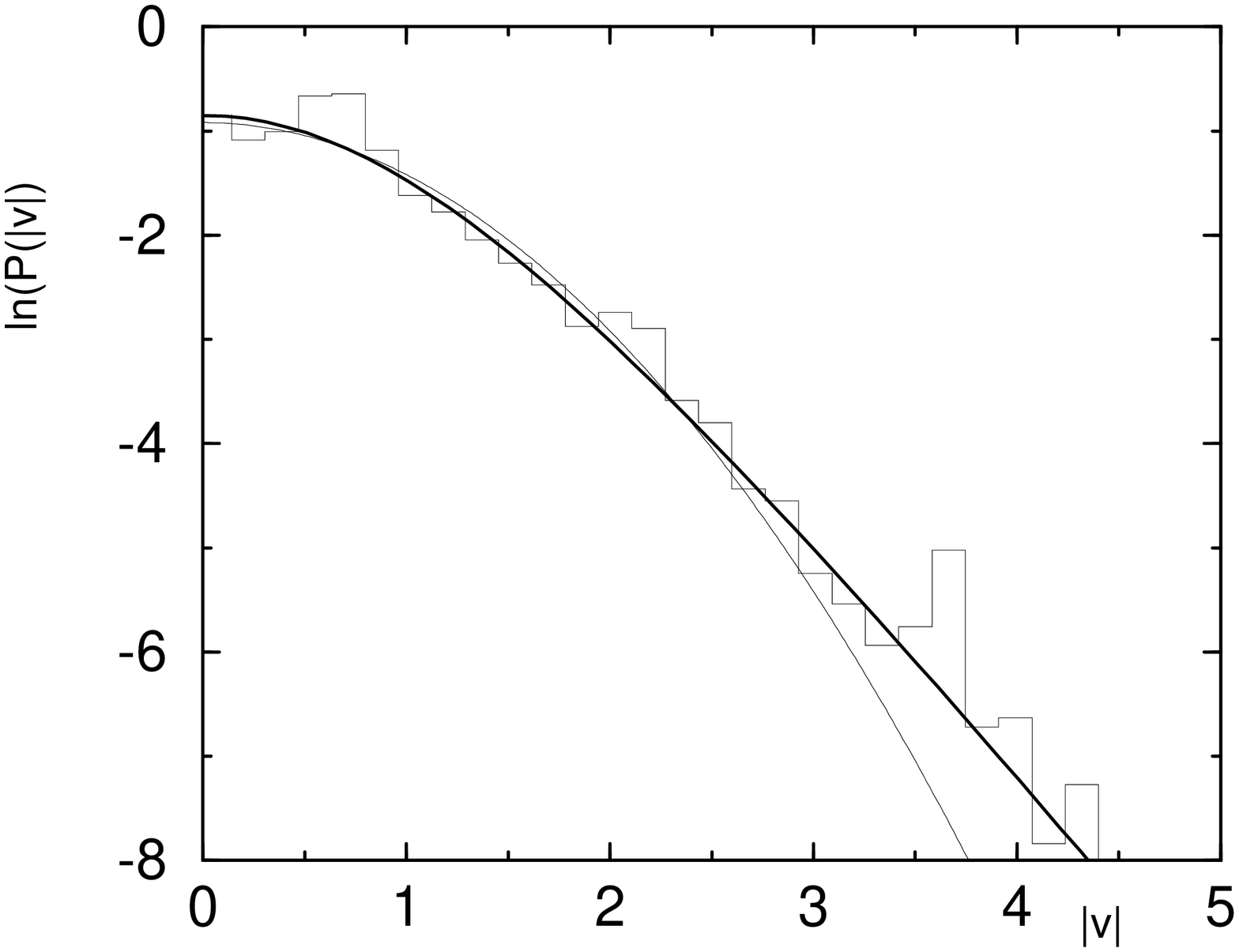,height=8cm,width=16cm,angle=270}
Fig.~9: Velocity distribution $P(v)$ for a fully connected hexagon.
$A=0$, and $\Lambda_i=\Lambda = 4.97$. The thick solid line correspond to
 ~(\ref{veldistr}) which describes the Poisson-GOE transition.
The thin solid line correspond to a Gaussian with the same mean and standard
deviation, with the numerical data.
\end{figure}

\begin{figure}
\psfig{figure=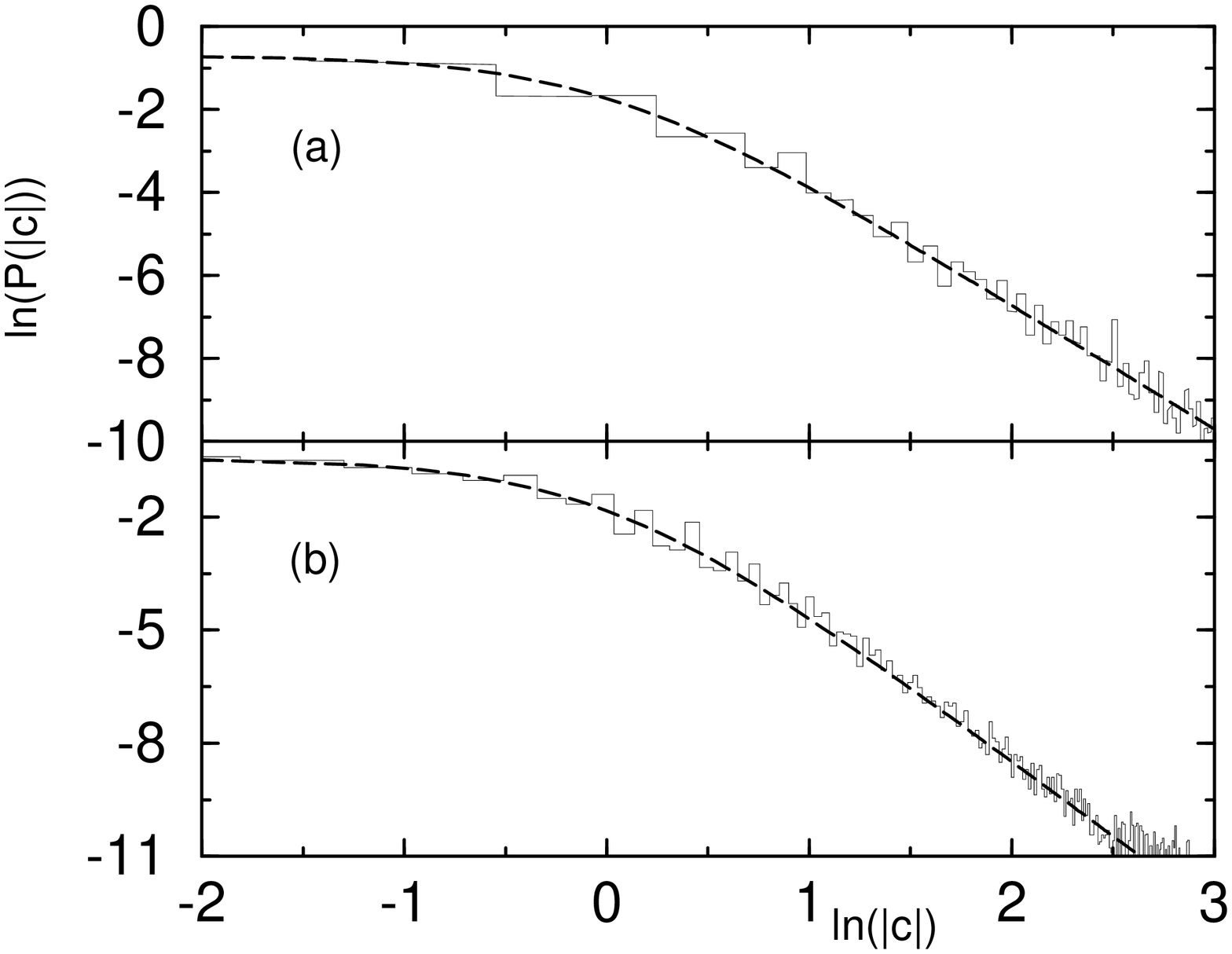,height=8cm,width=16cm,angle=270}
Fig.~10: Curvature distribution $P(c)$ for the fully connected hexagon. The
 thick
dashed line correspond to  ~(\ref{curvdistr}) with $\beta$ defined from the
symmetry of the system.\\
(a) $A=0$ and $\Lambda_i = 0$ (Neumann boundary conditions). The parameter
 $\beta = 1$.\\
(b) $A = 20$ and $\Lambda_i = 0$. The parameter $\beta=2$;\\
\end{figure}

\begin{figure}
\psfig{figure=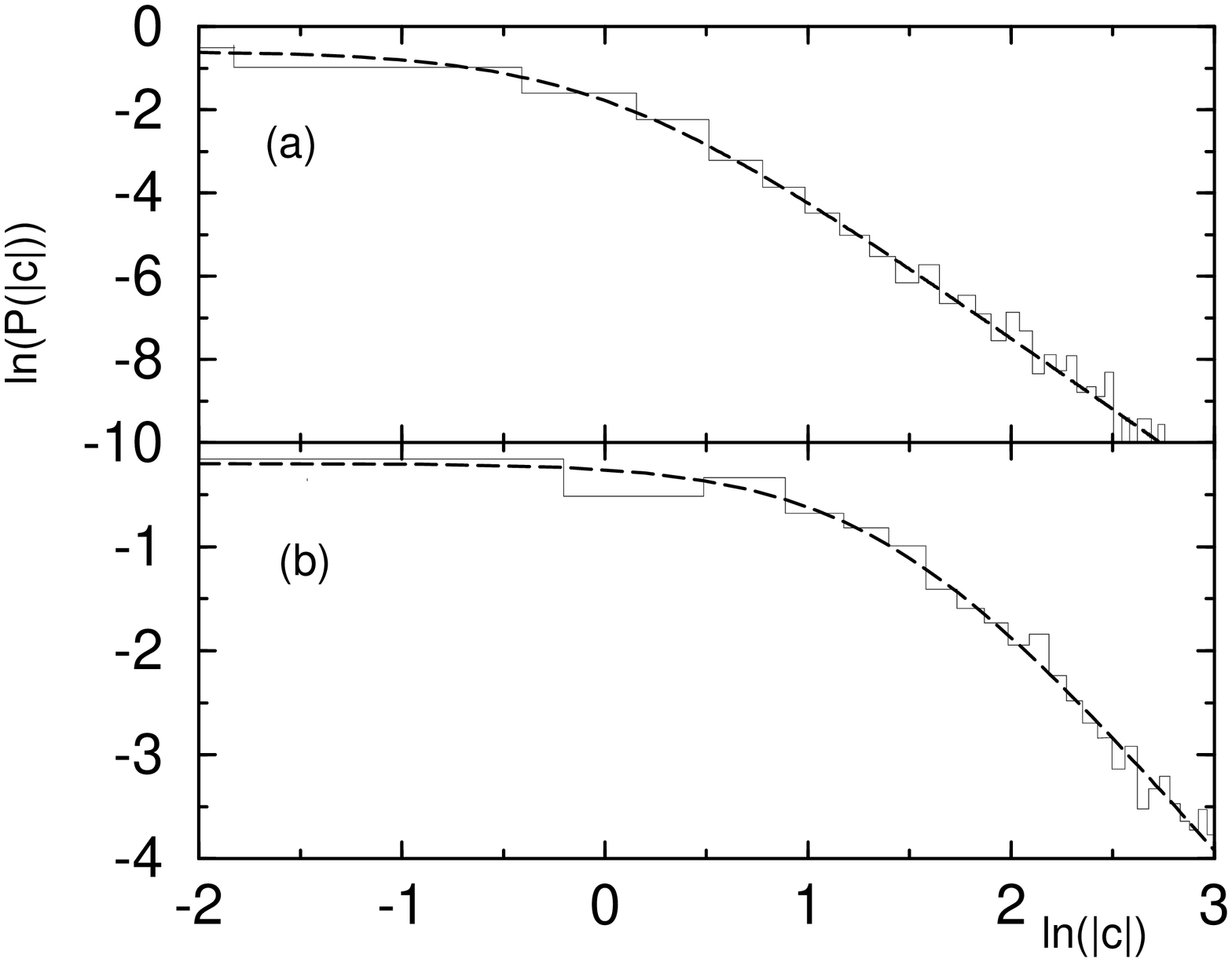,height=8cm,width=16cm,angle=270}
Fig.~11: Curvature distribution $P(c)$ for the fully connected hexagon. The
 thick
dashed line correspond to  ~(\ref{curvdistr}) with $\beta$ defined from the
symmetry of the system (fitting parameter).\\
(a) $A= 2$ and $\Lambda_i = 0$. The fitting parameter is $\beta = 1.4\pm
0.1$;\\
(b) $A=0$ and $\Lambda_i = \Lambda = 4.97$. The statistics was generated as
 explained
in the text. The fitting parameter is $\beta=0.28 \pm 0.01$.
\end{figure}

\begin{figure}
\psfig{figure=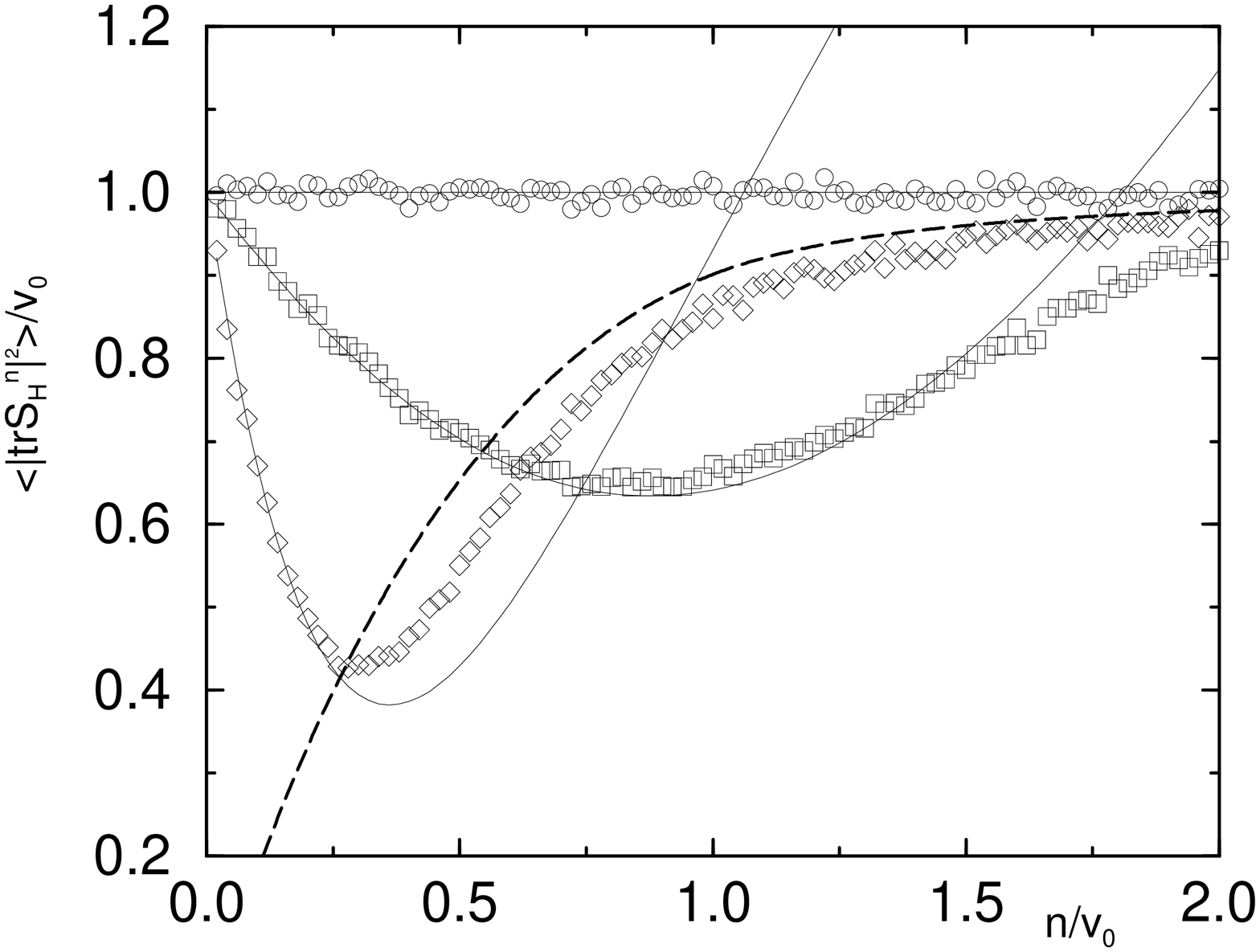,height=8cm,width=16cm,angle=270}
Fig.~12: The phase-shift form factor for a Hydra with $v_0=50$ and
 $\Lambda_0=2000$
($\circ$), $\Lambda_0=2$ ($\Box$) and $\Lambda_0=0$ ($\Diamond$). The thin
solid
lines correspond to the theoretical expectation (\ref{starfinal}), while the
 thick
dashed line to the RMT prediction.
\end{figure}

\begin{figure}
\psfig{figure=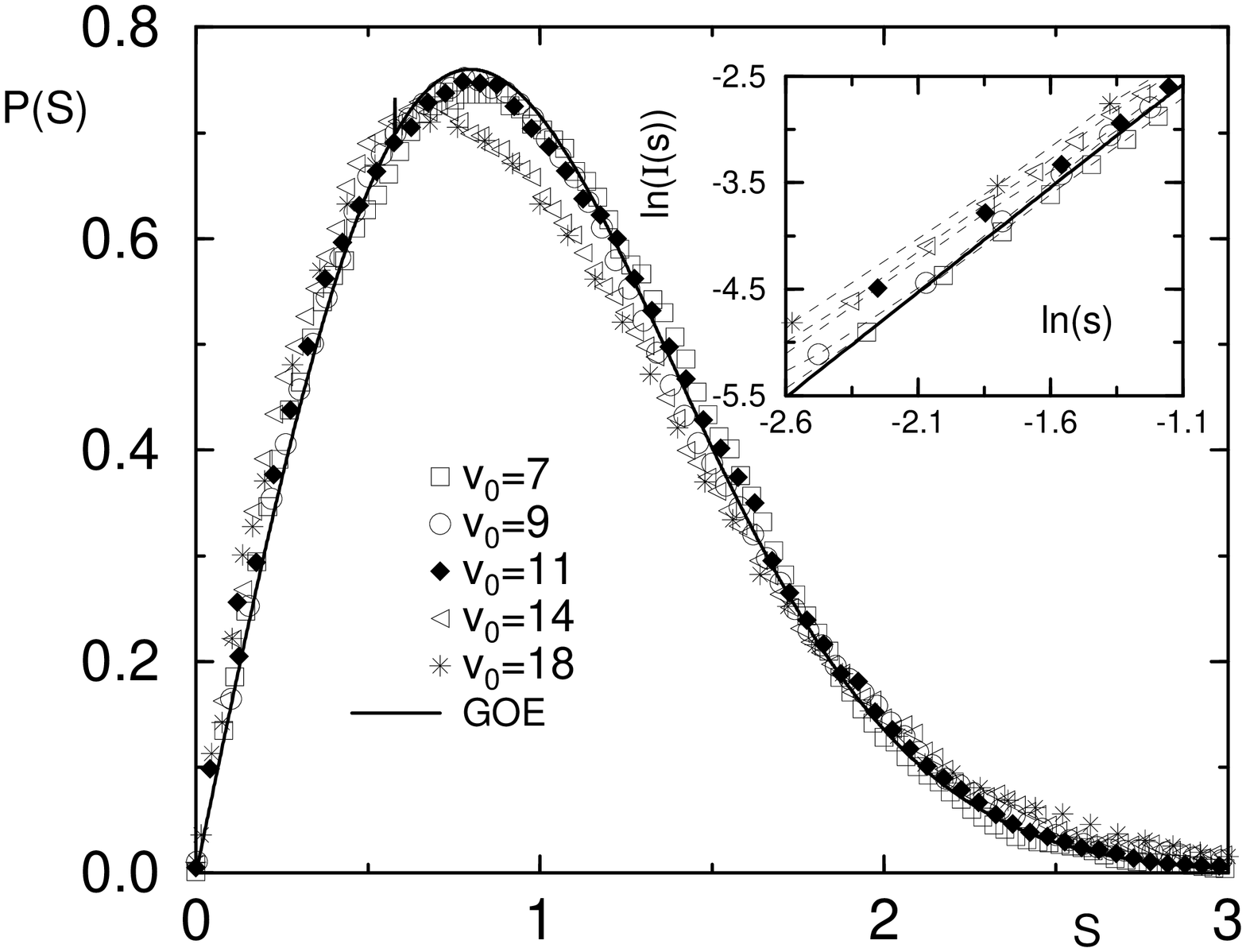,height=8cm,width=16cm,angle=270}
Fig.~13: Nearest level spacing distribution $P(s)$ for various valencies $v_0$
 of the Hydras and $\lambda_0 = 0$. In the inset we report the integrated level
 spacing $I(s)$ vs. $s$, for small spacings (the dashed lines are guides
for the
eye). Deviations from RMT (thick solid line) are clearly observed.
\end{figure}

\begin{figure}
\psfig{figure=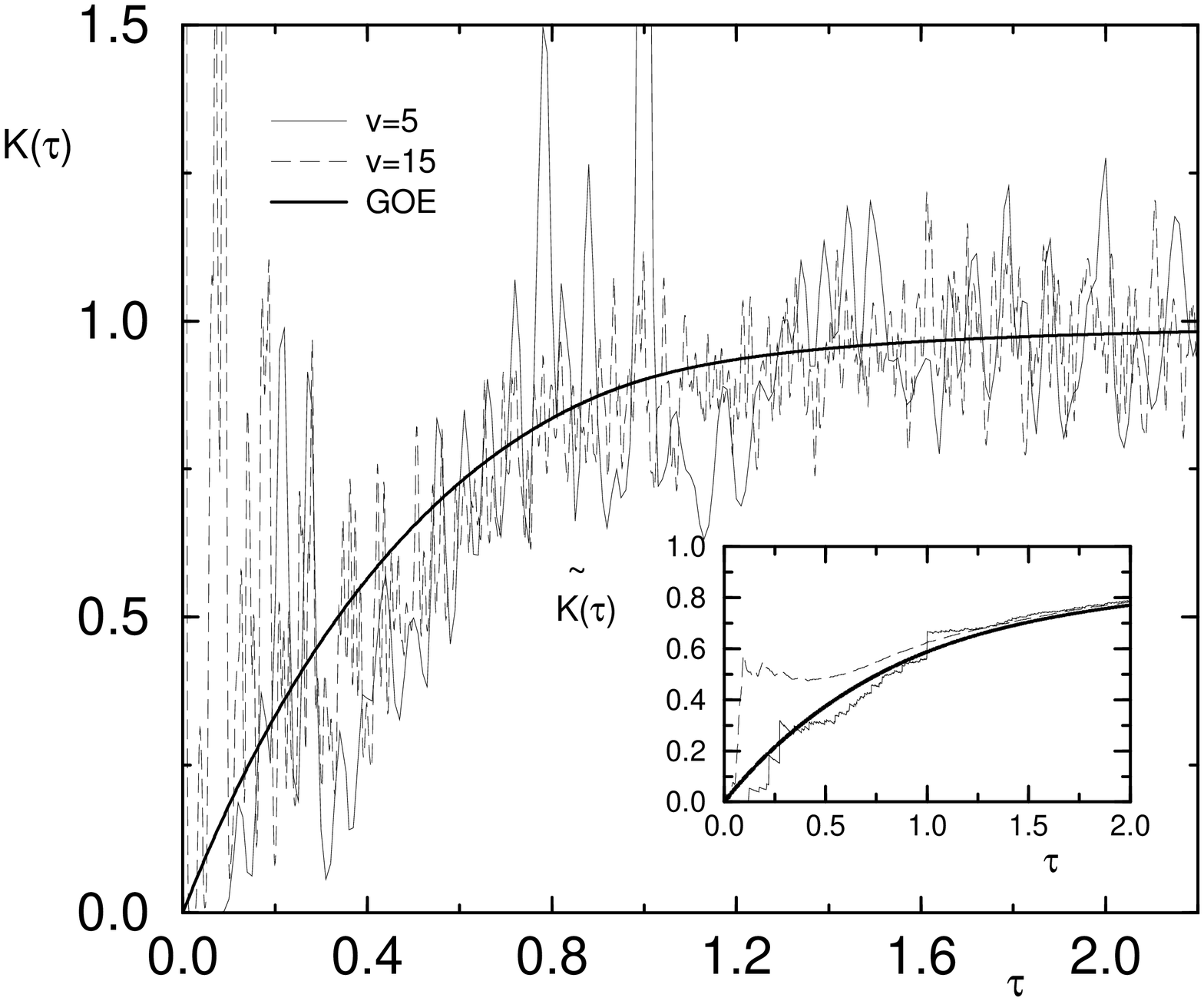,height=8cm,width=16cm,angle=270}
Fig.~14: The two-point form factor $K(\tau )$ for Hydras with $v_0 = 5$ (thin
 solid
line) and $v_0=15$ (thin dashed line). The thick solid line correspond to the
 RMT
expectations. In the inset we show the integrated form factor ${\widetilde
 K}(\tau)$.
\end{figure}

\begin{figure}
\psfig{figure=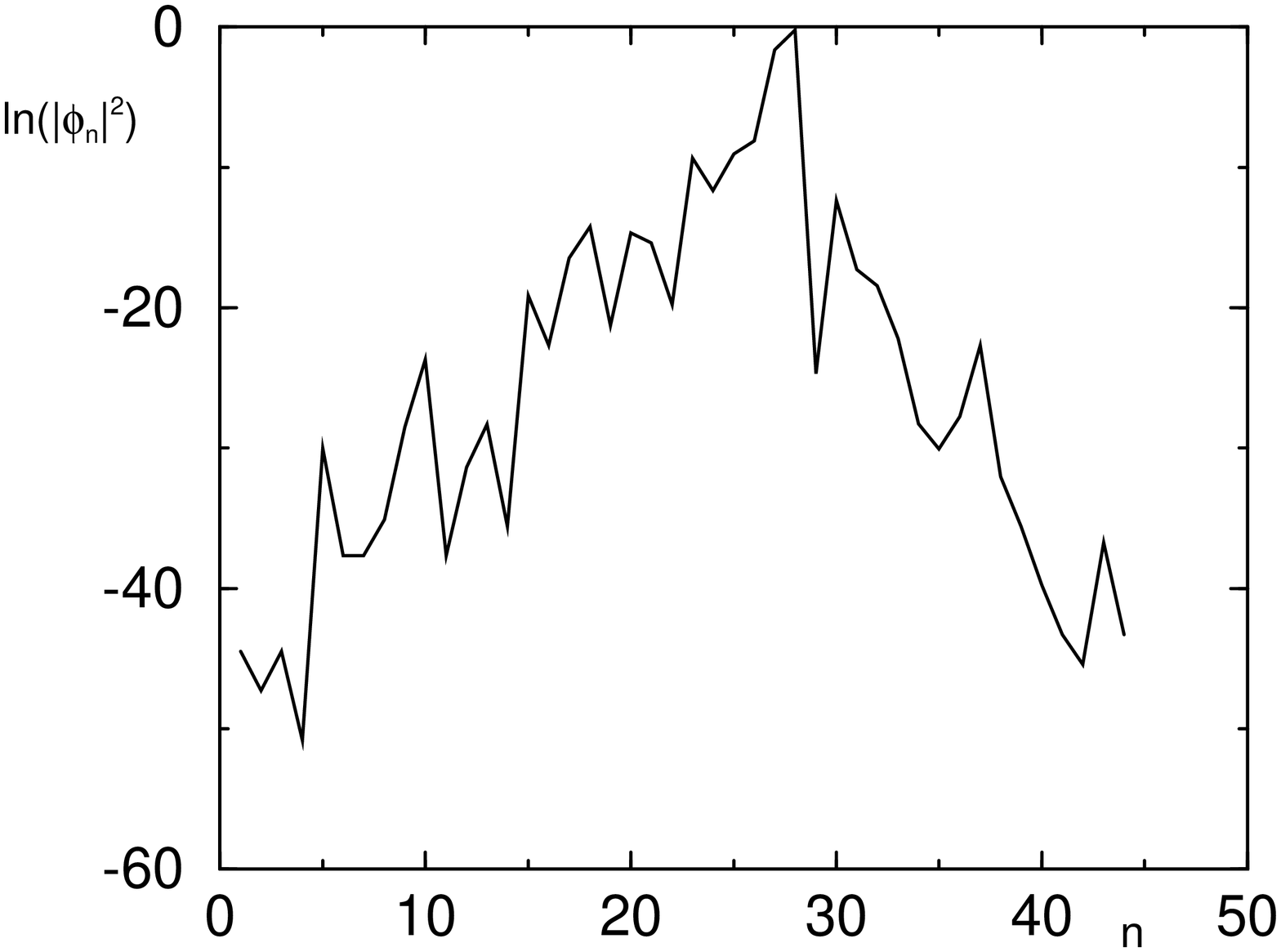,height=8cm,width=16cm,angle=270}
Fig.~15: A typical exponentially localized eigenstate of a graph with $V=44$
 and loops at each vertex.
\end{figure}

\begin{figure}
\psfig{figure=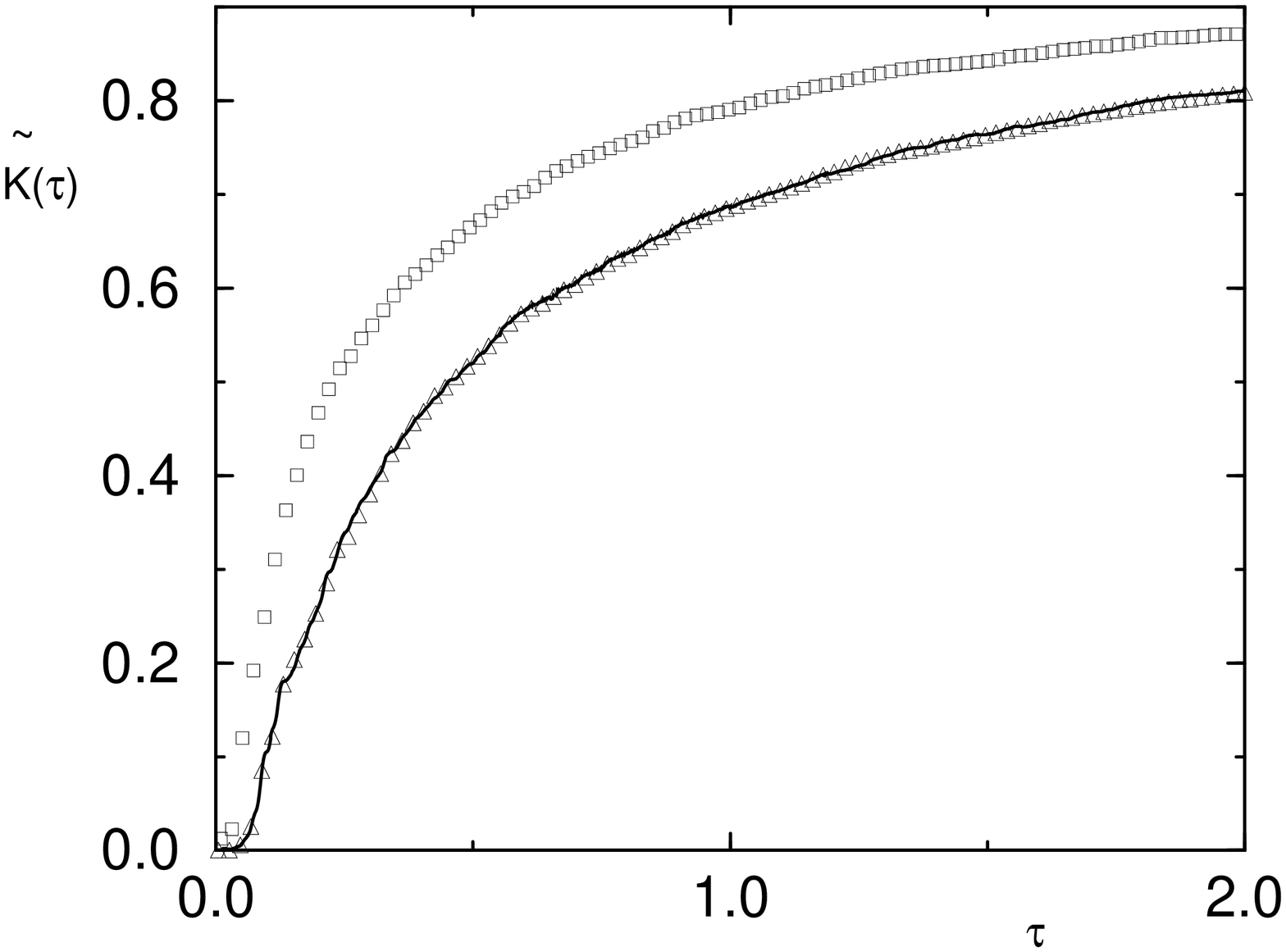,height=8cm,width=16cm,angle=270}
Fig.~16: The integrated form factor ${\widetilde K}(\tau)$ for graphs with
$V=11$ ($\triangle$) and $V=22$ ($\Box$) and loops at each vertex. The thick
 solid
line correspond to the graph with $V=22$ after rescaling $\tau$ by a factor
of two.
\end{figure}

\begin{figure}
\psfig{figure=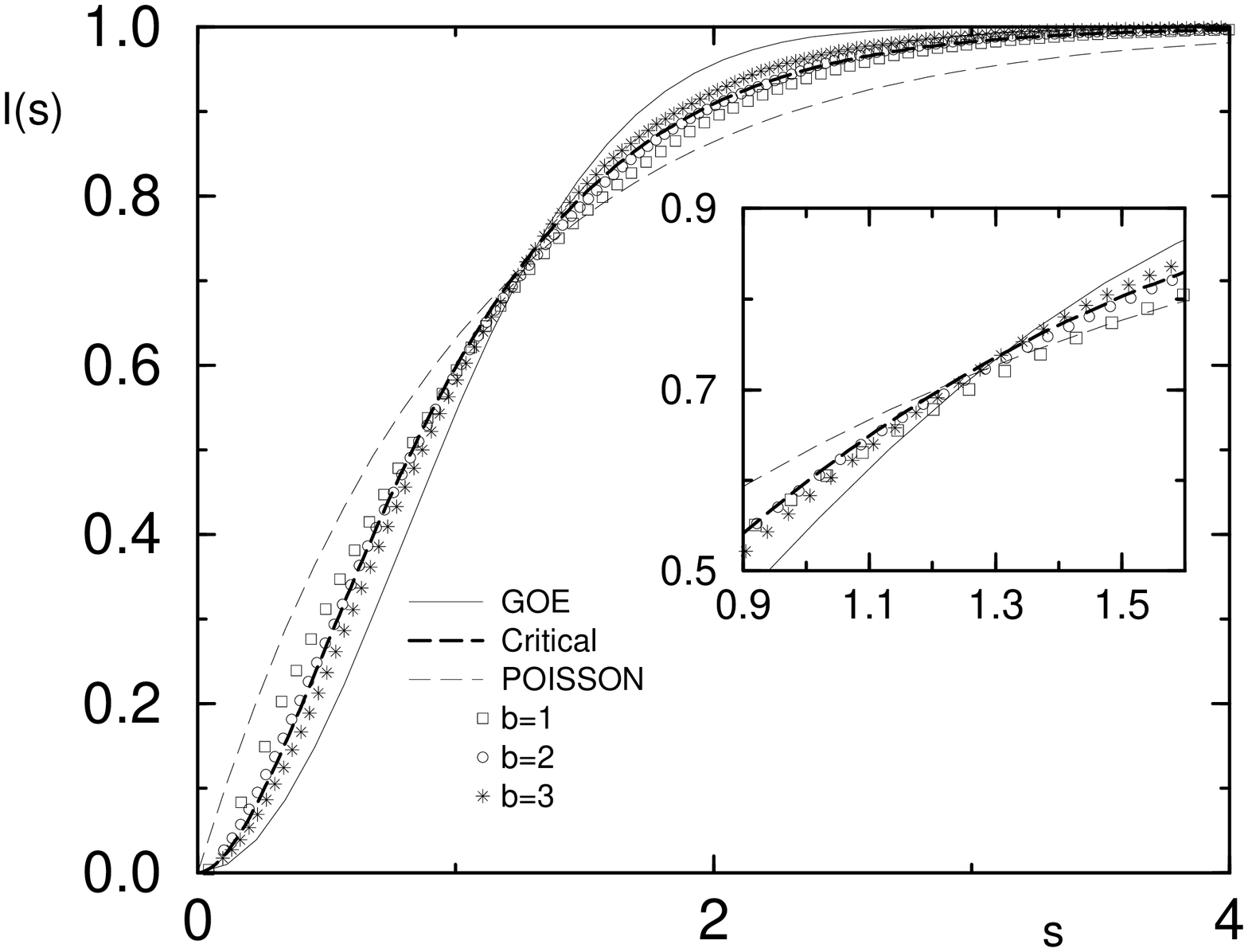,height=8cm,width=16cm,angle=270}
Fig.~17: Integrated nearest neighbors distribution $I(s)$ for a graph with
loops
 at
each vertex. The number of vertices is $V=22$. We consider various
 connectivities;
$b=1$ ($\Box$), $b=2$ ($\circ$), $b=3$ ($\ast$). The critical distribution
 (\ref{bog1})
is also shown (thick dashed line) together with the GOE (thin solid line) and
Poisson (thin dashed line) expressions.
\end{figure}

\begin{figure}
\psfig{figure=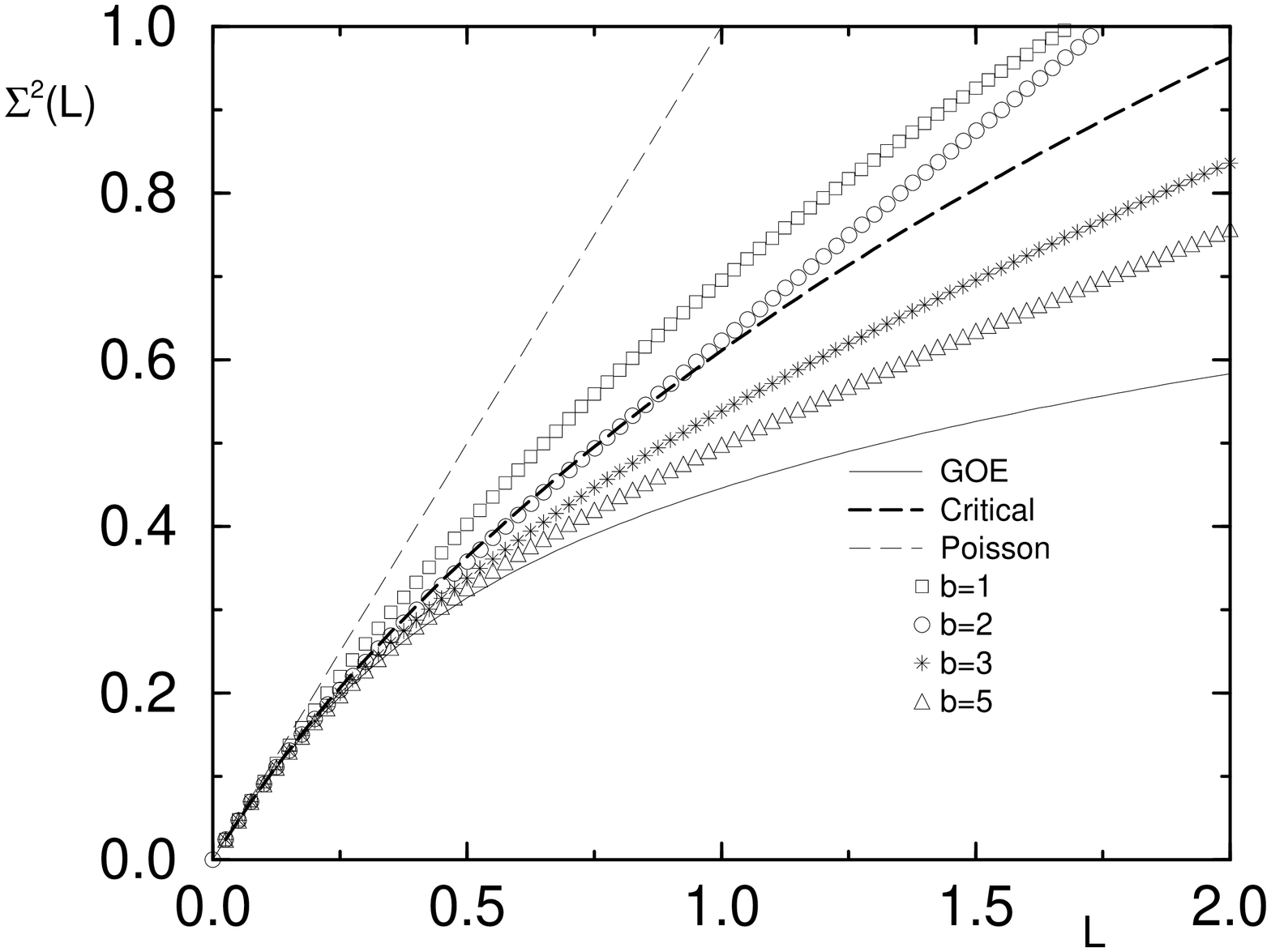,height=8cm,width=16cm,angle=270}
Fig.~18: The number variance $\Sigma^2(L)$ for the graph of Fig.~17 and various
connectivities; $b=1$ ($\Box$), $b=2$ ($\circ$), $b=3$ ($\ast$), $b=5$
 ($\triangle$).
The critical distribution (\ref{bogo2}) is shown by the thick dashed line,
 together
with the GOE (thin solid line) and Poisson (thin dashed line) results.
\end{figure}

\end{document}